\documentclass[times,sort&compress,3p]{elsarticle}
\journal{Journal of Multivariate Analysis}
\usepackage[labelfont=bf]{caption}

\usepackage{amsmath,amsfonts,amssymb,amsthm,booktabs,xcolor}
\usepackage{multirow,float,epsfig,graphicx,hyperref,url}
\theoremstyle{plain}
\theoremstyle{plain}
\newtheorem{theorem}{Theorem}

\newtheorem{proposition}{Proposition}
\newtheorem{lemma}{Lemma}

\theoremstyle{definition}

\newtheorem{remark}{Remark}
\newtheorem{example}{Example}
\newtheorem{condition}{Condition}
\newcommand{\ta}[1]{#1} 
\newcommand{\tO}[1]{#1} 
\newcommand{\td}[1]{} 

\def\av{\boldsymbol a}
\def\hv{\boldsymbol h}
\def\xv{\boldsymbol x}
\def\tv{\boldsymbol t}
\def\sv{\boldsymbol s}
\def\Uv{\boldsymbol U}
\def\Vv{\boldsymbol V}
\def\Wv{\boldsymbol W}
\def\Xv{\boldsymbol X}
\def\Yv{\boldsymbol Y}
\newcommand{\Sigmav}{\mbox{\boldmath{$\Sigma$}}}
\def\Xbf{\mathbf X}
\def\Ybf{\mathbf Y}

\def\1v{\mathbf 1}
\def\0v{\mathbf 0}

\newcommand\indep{\protect\mathpalette{\protect\independenT}{\perp}}\def\independenT#1#2{\mathrel{\rlap{$#1#2$}\mkern2mu{#1#2}}}

\newcommand{\argmin}{\operatornamewithlimits{argmin}}
\newcommand{\argmax}{\operatornamewithlimits{argmax}}

\begin{document}

\begin{frontmatter}

\title{Distribution-free and model-free multivariate feature screening via multivariate rank distance correlation}

\author[1]{Shaofei Zhao}
\author[1]{Guifang Fu\corref{mycorrespondingauthor}}

\address[1]{Department of Mathematical Sciences, Binghamton University, Vestal, NY 13850}

\cortext[mycorrespondingauthor]{Corresponding author. Email address: \url{gfu@binghamton.edu}}

\begin{abstract}
Feature screening approaches are effective in selecting active features from data with ultrahigh dimensionality and increasing complexity; however, many existing feature screening approaches are either restricted to a univariate response or rely on some distribution or model assumptions. In this article, we propose a sure independence screening approach based on the multivariate rank distance correlation (MrDc-SIS). The MrDc-SIS achieves multiple desirable properties such as being distribution-free, completely nonparametric, scale-free and robust for outliers or heavy tails. Moreover, the MrDc-SIS can be used to screen either univariate or multivariate responses and either one dimensional or multi-dimensional predictors. We establish the theoretical sure screening and rank consistency properties of the MrDc-SIS approach under a mild condition by lifting previous assumptions about the finite moments. Simulation studies demonstrate that MrDc-SIS outperforms eight other closely relevant approaches under some settings. We also apply the MrDc-SIS approach to a multi-omics ovarian carcinoma data downloaded from The Cancer Genome Atlas (TCGA).
\end{abstract}


\end{frontmatter}

\section{Introduction\label{sec:1}}

\ta{The explosion of big data brings unprecedented dimension and complexity challenges in a wide variety of fields. As a result, the well-established variable selection approaches, such as} the LASSO \citep{Tib}, smoothly clipped absolute deviation (SCAD, \citep{scad}), Elastic net \citep{elastic} and Dantzig selector \citep{Candes}, may have their effectiveness and accuracy reduced for ultrahigh dimensional data analyses \citep{samworth}. 

The concept of ``ultrahigh dimension'' was first introduced in \citet{sis}, defined as $\log(p) = O(n^\xi)$ for some $\xi>0$, \tO{here $p$ stands for the number of predictor variables, and $n$ is the sample size. It} is also called non-polynomial dimensionality or NP-dimensionality. \ta{The feature screening approaches are effective in selecting active predictors from ultrahigh dimensional data with theoretical guarantees \citep{sis,Li, mi,dcisis, liu2015selective, zhong2020forward, huang2014feature, cui2015model, chu2016feature, nandy2021covariate, zhou2017model, zhong2016regularized, wu2018network}.} However, the majority of existing feature screening approaches either explicitly or implicitly required some distribution or model assumptions. \citet{sis} pioneered the first sure independence screening (SIS) approach and selected predictors based on the Pearson correlation coefficient between each predictor and a univariate response\tO{, but the SIS approach required the normality and linear regression model assumptions.} \citet{sirs} proposed the sure independent ranking and screening (SIRS) approach by utilizing an indicator function to discretize the original response and then calculating the association between the indicator function and each of the predictors. The SIRS improved the SIS approach by lifting the normality and linear model assumptions. However, SIRS  still required finite moment assumptions on predictors to obtain the sure screening property. \citet{Li} proposed the distance correlation based sure independence screening (DC-SIS) approach. The DC-SIS also lifted the normality and model assumptions. However, it still required that both response and predictor variables meet the sub-exponential tail bound, which may not always be satisfied if the data has heavy tails or complex structures.  

Recently, several robust feature selection approaches have been studied. For example, \citet{zhong2016regularized} proposed a robust feature screening method for a univariate response (DC-RoSIS) by applying the DC-SIS to  original predictors and the rank statistic of response. As a result, DC-RoSIS is robust in terms of response but may not be robust for predictors. In addition, DC-RoSIS focused on a univariate response. \citet{pan2018generic} proposed a BCor-SIS approach based on Ball correlation, which is robust with mild assumptions and can be applied to both multivariate and univariate responses. \citet{liu2020model} proposed another robust model-free and data-adaptive feature selection method named PC-Screen, where they utilized projection correlation to measure the dependence and knockoff features to determine the threshold. The PC-Screen is also robust and can be applied to both multivariate and univariate responses, but its computation cost is relatively higher than other approaches. \citet{li2012robust} proposed a robust rank correlation screening (RRCS) based on Kendall's $\tau$. As commented by \citet{guo2022stable}, Kendall's $\tau$ excels in detecting monotone relationship but may not be effective in detecting other relationships between two variables. Moreover, the RRCS can not be directly applied to multivariate data. \citet{guo2022stable} invented a stable correlation (SC) as a new dependence score based on a new weight function of distance correlation and further extended it to a feature screening field (SC-SIS). Although the SC-SIS is robust and can be applied to multivariate predictors and responses, it requires to tune an extra parameter $a$ in the calculation of SC.

In this article, we propose a sure independence screening approach based on the multivariate rank distance correlation (\citep{rdc}), and refer it as MrDc-SIS. \citet{dc} systematically studied the theoretical property of distance correlation. \citet{Li} introduced the distance correlation into the feature screening field and proposed the DC-SIS. \citet{rdc} studied the theoretical property of multivariate rank distance correlation (MrDc) and in this article we further extend it into the feature screening field and propose the MrDc-SIS approach. The MrDc-SIS achieves multiple agreeable properties and significantly expands the capability of well-established extant screening approaches to better overcome the challenges associated with messy data.

Specifically, the proposed MrDc-SIS have the following good properties: 1) It achieves completely model-free and nonparametric properties, and is flexible for both linear and nonlinear relationships. Since the underlying true model is actually unknown in practice, it avoids inaccuracies caused by model misspecification. 2) It is robust, scale-invariant and distribution-free without requiring a normality assumption, which greatly expands its wide application scope in heavy-tailed data. Here the distribution-free means no matter what the original distribution of the data may be, it will always be transformed to a unit hypercube, thus the original distribution will not affect the performance of MrDc, see \citet[Proposition 2.2]{rdc} for more details. 3) The sure screening consistency property can be proven with the minimal condition compared with other existing feature screening methods. The sure screening consistency asymptotically guarantees that all true predictors are selected with probability approaching 1 as the sample size increases to $\infty$. The only condition it has on the variable is absolute continuity without restrictions on the moments of the underlying distributions. This relaxation of MrDc-SIS improves the selection success rate for heavy-tailed data such as $t$ and Pareto distributed data. 4) Unlike many feature screening approaches, MrDc-SIS is feasible for either a univariate or multivariate response, and either one dimensional or multi-dimensional predictors. Moreover, in the calculation of the MrDc between predictors and responses, no tuning parameter involved.

We perform three simulation studies with various difficulty levels. In simulation 1, we design a univariate response and compare MrDc-SIS with eight other relevant approaches, including SIS (\citet{sis}), SIRS (\citet{sirs}), RRCS (\citet{li2012robust}), SC-SIS (\citet{guo2022stable}), PC-Screen (\citet{liu2020model}), BCor-SIS (\citet{pan2018generic}), DC-SIS (\citet{Li}) and DC-RoSIS (\citet{zhong2016regularized}). In simulations 2-3, we focus on multivariate responses and compare MrDc-SIS with only four approaches that are feasible for multivariate responses, which are SC-SIS, PC-Screen, Bcor-SIS, and DC-SIS. Simulation studies demonstrate the robustness and outperformance of the MrDc-SIS under some settings, especially in one-tailed and fat-tailed data, like Pareto distributions. We also apply the MrDc-SIS approach to ovarian carcinoma (OV) downloaded from The Cancer Genome Atlas (TCGA). The data was collected from multiple platforms, such as genome and epigenome, for the same patient (it is called ``multi-omics'' data). The modeling aim is to detect active genes associated with OV and in the long run provide theoretical guidance for the diagnosis and prognosis of the disease. The exploratory data analysis shows that this multi-omics data has extremely long tails and large ranges of scales (a range from 0 to 50,000 as an example); nonlinear associations, and complex structures. 

The remainder sections are organized as follows: In Section 2, we elaborate on the details and theoretical properties of MrDc-SIS. This is followed by an assessment of the finite sample performance of MrDc-SIS via simulation studies in Section 3. In section 4 we implement the MrDc-SIS to multiple platforms of the TCGA data. In Section 5, we discuss the method and give our conclusion. Finally, the proof of the main theorem is given in the appendix.

\section{Methods}
\label{sec:2}
\subsection{Multivariate rank}

 \citet{rdc} initiated a distribution-free and model-free dependence measure by integrating multivariate rank with distance correlation and they used low-discrepancy sequences to map the original data to a unit hypercube. Let \ta{$\mathbb{R}^d$ denote the original $d$-dimensional space and $[0,1]^d$ denote the $d$-dimensional unit hypercube that the original data is mapped into by the multivariate rank process.} Let $\mathcal{P}(\mathbb{R}^d)$ denote the families of all probability distributions on $\mathbb{R}^d$, and $\mathcal{P}_{ac}(\mathbb{R}^d)$ denote the families of Lebesgue absolutely continuous probability measures on $\mathbb{R}^d$. Let $\mathcal{U}^d$ denote the uniform distribution on $[0,1]^d$ and $C_n$ stands for the set of all permutations of $\{1, \dots, n\}$. The multivariate rank is implemented through a measure transportation, or optimal transportation. It is a problem of finding a ``nice'' function $G:\mathbb{R}^d\to\mathbb{R}^d$ such that $G$ maps a given measure $\mu\in\mathcal{P}(\mathbb{R}^d)$ to $\nu\in\mathcal{P}(\mathbb{R}^d)$, written as $G\#\mu = \nu$. It means that $G(\Xbf)\sim\nu$ where $\Xbf\sim\mu$ \citep{shi, rdc}.
\begin{proposition}[McCann's Theorem \citep{mcc}]
Suppose $\mu,\nu\in\mathcal{P}_{ac}(\mathbb{R}^d)$, then there exists a function $R(\cdot)$, which is the gradient of an (extended) real-valued d-variate convex function, such that $R\#\mu=\nu$. $R$ is unique $\mu$ $a.e.$. Moreover, if $\mu$ and $\nu$ have finite second moments, $R(\cdot)$ is also the solution to the Monge's problem:
\begin{equation*}
\inf_F\int\Vert \Xbf - G(\Xbf)\Vert^2 d\mu(\Xbf)\; \textup{ subject to }\; G\#\mu = \nu.
\end{equation*}
\end{proposition}

Based on \ta{the Proposition 1}, \tO{if we let $\nu = \mathcal{U}^d$ and give a measure $\mu\in\mathcal{P}_{ac}(\mathbb{R}^d)$, \ta{then there exists a rank function}, $R(\cdot)$, such that $R\#\mu = \nu$,} and this is unique up to measure zero sets with respect to $\mu$. However, in practice we don't really know the distribution $\mu$, instead our only knowledge about $\mu$ is obtained from  \ta{data observations,} $\Xbf_1,\dots,\Xbf_n \overset{\text{i.i.d}}{\sim} \mu\in\mathcal{P}_{ac}(\mathbb{R}^d )$. If $\{X_1,\dots.X_n\}$ are samples of a univariate random variable $X$ \ta{(i.e., $d=1$)}, it is easy to get a good discrete approximation of $ \mathcal{U}^1$  by simply sorting $\{X_i\}_{i=1}^n$ and assigning $\{i/n\}_{i=1}^n$ to the sorted points. However, if $\{\Xbf_i\}_{i=1}^n$ are \ta{samples of a multivariate vector} with higher dimensions ($d>1$), \ta{we utilize the low-discrepancy sequences to approximate $\mathcal{U}^d$}.

\begin{figure}[h!]
\centering
\begin{minipage}{.48\textwidth}
\includegraphics[width=\textwidth]{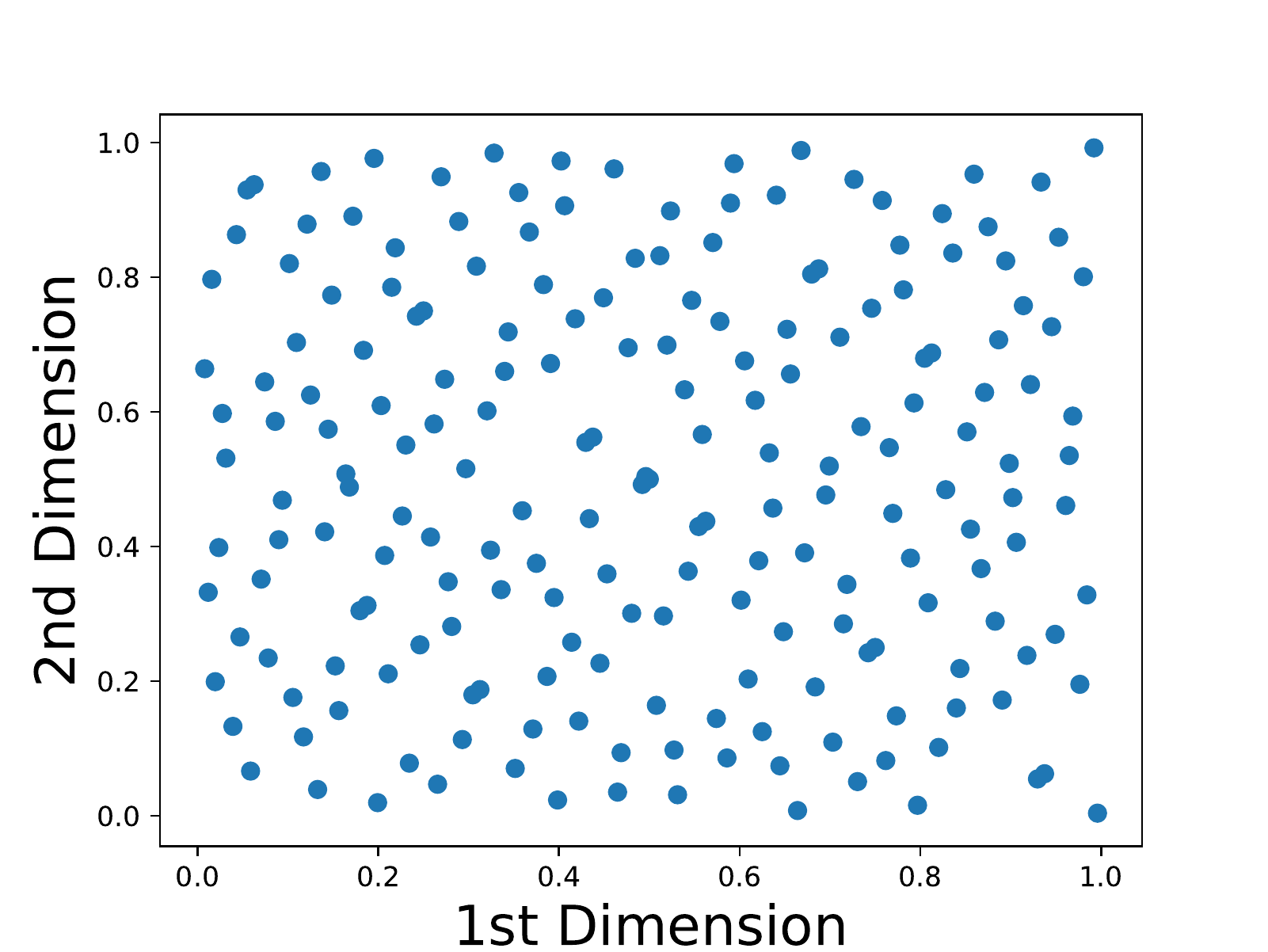}
\end{minipage}
\hfil
\begin{minipage}{.48\textwidth}
\includegraphics[width=\textwidth]{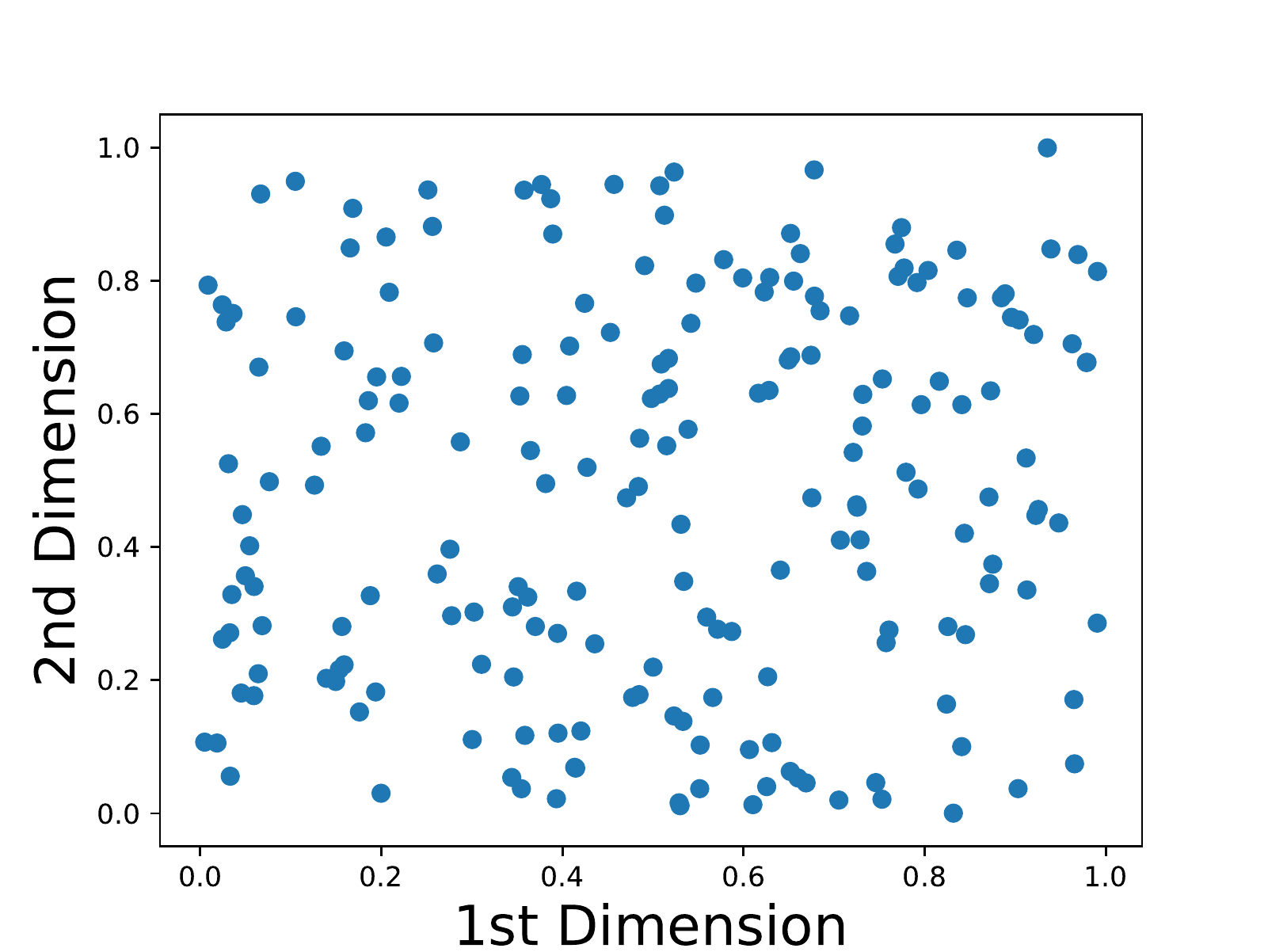}
\end{minipage}
\caption{\ta{Plots of the 1st and 2nd dimension for the Sobol' sequence (left panel) and random sequence (right panel) for a sample with 200 observations.} 
The Sobol' sequence looks more uniform or less ``discrepancy'' compared with the random generator.}
\label{discrepancy}
\end{figure}

Low-discrepancy sequence, also known as the quasi monte-carlo (QMC) sequence, is to construct a fixed-points set with low ``discrepancy'' \citep{josef}. The property of ``low-discrepancy'' means that the proportion of points in the sequence falling into an arbitrary set $B$ is close to the proportional of the measure of $B$. In other words, even if we have the same number of points in both sequences, the low-discrepancy sequences may be more equally distributed on $[0,1]^d$ than random sequences. Among these low-discrepancy sequences, Halton sequence \citep{halton}, Sobol' sequence \citep{sobol}, Niederreiter sequence (also known as (t,m,s)-nets; \citep{nied}), and their extensions (scrambled, truncated, etc) have been thoroughly studied.  \citet{sobol16384} explored the Sobol' sequence up to 16,384 dimensions; and \citet{joe} successfully constructed Sobol' sequence for \ta{21,201-dimensional data, which was an extremely large experiment. Theoretically speaking, it is feasible to extend low-discrepancy sequences to dimensions that are even higher than those explored in the extant literature}, but it may be difficult to check the low-discrepancy properties and to avoid collinearity.  \ta{Fig. \ref{discrepancy} illustrates a comparison of Sobol' sequence and a random sequence, from which we can see that the Sobol' sequence distributes more evenly and more uniformly than the random sequence.} 

Let $\mathcal{D}_n^{\Xbf}=\{ \Xbf_1^d,\dots,\Xbf_n^d \}$ be the observed data, where \tO{each data point has $d$-dimensions}. Let $\mathcal{H}_n^d:=\{\hv_1^d,\dots,\hv_n^d\}$ denote a sample of $d$-variate vector after multivariate rank map (we use Sobol' sequences for $d\geqslant 2$; and use $\{i/n\}_{i = 1}^n$ for $d = 1$). Let $\mu_n^{\Xbf}:=\sum_{i = 1}^{n}\delta_{\Xbf_i^d}/n$ and $\nu_n:=\sum_{i = 1}^{n}\delta_{h_i^d}/n$ be the empirical distributions on $\mathcal{D}_n^{\Xbf}$ and $\mathcal{H}_n^d$ respectively, where $\delta$ represents the Dirac measure. Then the empirical rank is defined as the optimal transport map which transports $\mu_n^{\Xbf}$ to $\nu_n$, that is,
\begin{equation*}
  \hat{R}_n = \argmin_G\int\Vert \Xbf-G(\Xbf)\Vert^2 d\mu_n^{\Xbf}(\Xbf)\ \text{subject to}\ G\#\mu_n^{\Xbf} = \nu_n.
\end{equation*}

It is equivalent to
\begin{equation*}
  \hat{\sigma}_n:=\argmin_{\sigma\in C_n}\sum_{i = 1}^{n}\Vert \Xbf_i^d-\hv_{\sigma(i)}^d\Vert^2 = \argmax_{\sigma\in C_n}\sum_{i = 1}^{n}\langle \Xbf_i^d,\hv_{\sigma(i)}^d\rangle.
\end{equation*}
Finally, the empirical rank map is obtained as
\begin{equation}
  \hat{R}_n(\Xbf_i^d) = \hv_{\hat{\sigma}_n(i)}^d,\  i \in\{ 1,\dots,n\}.
  \label{Rhat}
\end{equation}

Equation (\ref{Rhat}), i.e., finding the argmin, is an assignment problem. We utilized the modified Hungarian algorithm \citep{edm, tom} that are implemented as standard solvers in most platforms (e.g., \texttt{scipy.optimize.linear\_sum\_assignment} in \textsf{Python}).

\subsection{Multivariate rank distance correlation}

Let $\Xbf$ and $\Ybf$ be two multivariate vectors with dimensions $s_1$ and $s_2$, respectively. The optimal rank map $R^\Xbf(\Xbf)$ transforms $\Xbf$ into $\mathcal{H}^{s_1}$, and the optimal rank map $R^{\Ybf}(\Ybf)$ transforms $\Ybf$ into $\mathcal{H}^{s_2}$. Let $\phi_{R^\Xbf}(\tv)$, $\phi_{R^\Ybf}(\sv)$ denote the individual characteristic functions, and $\phi_{R^\Xbf,R^\Ybf}(\tv,\sv)$ denotes the joint characteristic function of $R^\Xbf(\Xbf)$ and $R^\Ybf(\Ybf)$. Then the multivariate rank distance covariance (MrDcov) between $\Xbf$ and $\Ybf$ is defined as
\begin{equation*}
  \mathrm{MrDcov}^2(\Xbf,\Ybf) = \int_{\mathbb{R}^{s_1+s_2}}\Vert\phi_{R^\Xbf,R^\Ybf}(\tv,\sv)-\phi_{R^\Xbf}(\tv)\phi_{R^\Ybf}(\sv)\Vert^2 w(\tv,\sv)d\tv d\sv,
\end{equation*}
where
\begin{equation*}
  w(\tv,\sv) = (c_{s_1}c_{s_2}\Vert\tv\Vert_{s_1}^{1+s_1}\Vert\sv\Vert_{s_2}^{1+s_2})^{-1},
\end{equation*}
with $c_d = \pi^{(1+d)/2}/\Gamma((1+d)/2)$ as a weight function. Throughout this paper, $\Vert\av\Vert_{d}$ represents the Euclidean norm of $\av\in \mathbb{R}^d$, and $\Vert\phi\Vert^2 = \phi\bar{\phi}$ for complex-valued function $\phi$ with $\bar{\phi}$ being its conjugate. 

Accordingly, the multivariate rank distance correlation is defined as
\begin{equation*}
  \mathrm{MrDc}(\Xbf,\Ybf) = \frac{\mathrm{MrDcov}(\Xbf,\Ybf)}{\sqrt{\mathrm{MrDcov}(\Xbf,\Xbf)\mathrm{MrDcov}(\Ybf,\Ybf)}}.
\end{equation*}
Compared to the original distance correlation, multivariate rank distance correlation has several remarkable advantages, which make it agreeable to construct a sure screening procedure based on MrDc:
\begin{itemize}
  \item It does not require the original data, $\Xbf$ and $\Ybf$, to have finite first moments, because $R^\Xbf(\Xbf)$ and $R^\Ybf(\Ybf)$ already lie in the unit hypercube.
  \item It does not require any distribution assumptions for $\Xbf$ and $\Ybf$, and the exponential tail bound condition that was required in \citet{Li} is automatically satisfied.
  \item Under a special case when $(\Xbf,\Ybf)$ follows a bivariate Gaussian distribution with mean vector $\0v$, variances 1 and correlation $\rho$, $\mathrm{MrDc}(X,Y)$ and the original distance correlation both increases when $|\rho|$ increases \citep[C.3]{rdc}. 
\end{itemize}

According to \citet{dc}, the multivariate rank distance covariance can be computed as,
\begin{equation*}
  \mathrm{MrDcov}^2(\Xbf,\Ybf) = \mathrm{MrS}_1+\mathrm{MrS}_2-2 \mathrm{MrS}_3,
\end{equation*}
where
\begin{equation*}
  \begin{split}
     \mathrm{MrS}_1 & = E[\Vert R^\Xbf(\Xbf)-R^\Xbf(\tilde{\Xbf})\Vert_{s_1}\Vert R^\Ybf(\Ybf)-R^\Ybf(\tilde{\Ybf})\Vert_{s_2}], \\
     \mathrm{MrS}_2 & = E[\Vert R^\Xbf(\Xbf)-R^\Xbf(\tilde{\Xbf})\Vert_{s_1})E(\Vert R^\Ybf(\Ybf)-R^\Ybf(\tilde{\Ybf})\Vert_{s_2}], \\
     \mathrm{MrS}_3 & = E[E[\Vert R^\Xbf(\Xbf)-R^\Xbf(\tilde{\Xbf})\Vert_{s_1}|R^\Xbf(\Xbf)]]E[E[\Vert R^\Ybf(\Ybf)-R^\Ybf(\tilde{\Ybf})\Vert_{s_2}|R^\Ybf(\Ybf)]].
  \end{split}
\end{equation*}
Here $(R^\Xbf(\tilde{\Xbf}),R^\Ybf(\tilde{\Ybf}))$ is an independent copy of $(R^\Xbf(\Xbf),R^\Ybf(\Ybf))$.

Specifically, the estimation process is as follows: Given the dataset $\{\Xbf_i, \Ybf_i\}_{i = 1}^n$ ($\Xbf_i$ is $s_1$ dimension and $\Ybf_i$ is $s_2$ dimension), \ta{we obtain $\{\hat {R}_n^\Xbf(\Xbf_i), \hat {R}_n^\Ybf(\Ybf_i)\}_{i = 1}^n$ from the multivariate rank in (\ref{Rhat}) using Sobol's sequence}. Then we estimate the $\mathrm{MrS}_1$, $\mathrm{MrS}_2$ and $\mathrm{MrS}_3$ as
\begin{equation*}
  \begin{split}
     \widehat{\mathrm{MrS}}_1 & = \frac{1}{n^2}\sum_{i=1}^{n}\sum_{j=1}^{n}\Vert\hat{R}_n^{\Xbf}(\Xbf_i)-\hat{R}_n^{\Xbf}(\Xbf_j)\Vert_{s_1}\Vert\hat{R}_n^{\Ybf}(\Ybf_i)
     -\hat{R}_n^{\Ybf}(\Ybf_j)\Vert_{s_2}, \\
     \widehat{\mathrm{MrS}}_2  & = \frac{1}{n^2}\sum_{i=1}^{n}\sum_{j=1}^{n}\Vert\hat{R}_n^{\Xbf}(\Xbf_i)-\hat{R}_n^{\Xbf}(\Xbf_j)\Vert_{s_1} \frac{1}{n^2}\sum_{i=1}^{n}\sum_{j=1}^{n}\Vert\hat{R}_n^{\Ybf}(\Ybf_i)-\hat{R}_n^{\Ybf}(\Ybf_j)\Vert_{s_2}, \\
     \widehat{\mathrm{MrS}}_3 & = \frac{1}{n^3}\sum_{i=1}^{n}\sum_{j=1}^{n}\sum_{l=1}^{n}\Vert\hat{R}_n^{\Xbf}(\Xbf_i)-\hat{R}_n^{\Xbf}(\Xbf_l)\Vert_{s_1}
     \Vert\hat{R}_n^{\Ybf}(\Ybf_j)-\hat{R}_n^{\Ybf}(\Ybf_{\ell})\Vert_{s_2}.
     \end{split}
\end{equation*}
Finally, the multivariate rank distance covariance and correlation can be estimated as
\begin{equation*}
    \begin{split}
     \widehat{\mathrm{MrDcov}}^2(\Xbf,\Ybf) & = \widehat{\mathrm{MrS}}_1 +\widehat{\mathrm{MrS}}_2-2\widehat{\mathrm{MrS}}_3, \\
     \vspace{6em}
     \widehat{\mathrm{MrDc}}(\Xbf,\Ybf) & =\frac{\widehat{\mathrm{MrDcov}}(\Xbf,\Ybf)}{\sqrt{\widehat{\mathrm{MrDcov}}(\Xbf,\Xbf)
     \widehat{\mathrm{MrDcov}}(\Ybf,\Ybf)}}.
  \end{split}
\end{equation*}

\subsection{Independence screening procedure based on multivariate rank distance correlation}
\tO{We propose a sure independence screening procedure based on the MrDc in this section. Let $\Ybf = (Y_1,\dots,Y_q)^\top$ ($q$ is the dimension of response and is fixed) be the response vector with support $\Psi_\Ybf$, $\Xbf = (X_1,\dots,X_p)^\top$ be the predictor vector and each predictor is $d$-dimensional. Let $F(\Ybf|\Xbf)$ be the conditional distribution function of $\Ybf$ given $\Xbf$.} In ultrahigh dimensional settings the number of predictors ($p$) may exceed the number of observations ($n$) exponentially, and only a small portion of predictors are truly relevant to the response (i.e., sparse structure). The predictors can accordingly be divided into two parts: active predictors which are truly related with the response and inactive predictors which are not related with the response. Define
\begin{equation*}
  \begin{split}
     \mathcal{D} & = \{j: F(\Ybf|\Xbf)\ \text{ functionally depends on }X_j\ \text{for some }\Ybf\in\Psi_\Ybf\}, \\
     \mathcal{I}  & =\{j: F(\Ybf|\Xbf)\ \text{does not functionally depend on }X_j \text{ for any }\Ybf\in\Psi_\Ybf\}.
  \end{split}
\end{equation*}
as the index sets of active and inactive predictors. Accordingly, $\xv_{\mathcal{D}} = \{X_j: j\in\mathcal{D}\}$ and $\xv_{\mathcal{I}} = \{X_j: j\in\mathcal{I}\}$ as active and inactive predictors sets. \ta{The aim of the MrDc-SIS approach} is to identify the index set $\mathcal{D}$ from all indices of the entire candidate pool. From the definition above, we can see $\Ybf\indep\xv_{\mathcal{I}}|\xv_{\mathcal{D}}$, where $\indep$ denotes statistical independence, so $\xv_{\mathcal{I}}$ are redundant when $\xv_{\mathcal{D}}$ are known.

Given the dataset, $\{\Xbf_i,\Ybf_i\}_{i = 1}^n$, define 
\begin{equation*}
  \omega_j = \mathrm{MrDc}^2(X_j,\Ybf), \text{ and } \hat{\omega}_j = \widehat{\mathrm{MrDc}}^2(X_j,\Ybf), \ j\in\{1,\dots,p\}.
\end{equation*}
\ta{Here $\omega_j$ is a dependence score that measures the association strength between each predictor $X_j$ and a multivariate response vector $\Ybf$, and $\hat{\omega}_j$ is the sample estimate of $\omega_j$, which can be used to rank the predictors from the most important to the least important.} The finally selected subset of active predictors is defined as
\begin{equation*}
  \hat{\mathcal{D}} = \{j: \hat{\omega}_j\geqslant cn^{-\kappa}, 1\leqslant j \leqslant p\},
\end{equation*}
where $c$ and $\kappa$ are pre-specified threshold values which will be defined in next subsection.

\subsection{Theoretical properties}
We impose the following minimum signal strength conditions.
\begin{condition} (Minimum signal strength)
\begin{enumerate}[(a)]
\item For some $c>0$ and $0\leqslant \kappa <1/2$, $\min_{j\in\mathcal{D}}\omega_j\geqslant 2cn^{-\kappa}$.

\item For some $c_2 >0 $ and $0\leqslant \kappa_2 <1/2$, $\min_{j\in\mathcal{D}}\omega_j - \max_{j \in \mathcal{I}}\omega_j \geqslant 2c_2n^{-\kappa_2}$.
\end{enumerate}
\end{condition}

\begin{remark}
Condition 1(a) assumes that the minimum dependence score in the active set should be greater than a positive value. This assumption is very common in the feature screening literature, for example \citet{Li} condition (C2), \citet{sis} condition 3, \citet{ifort} condition (C3), \citet{huang2014feature} condition (C2), and \citet{pan2018generic} condition (C1), among many others. This condition also makes more sense in practice because a so-called active predictor should at least have nonzero effects. Condition 1(b) is stronger than Condition 1(a), and it assumes that there is a gap of signal strength between active features and inactive features. This is the same as condition 1(b) in \citep{liu2020model} and condition (C2) in \citep{guo2022stable}, and it is a mild condition as we allow the gap to tend to 0 as $n\to\infty$.
\end{remark}
\color{black}
\begin{theorem}(Sure screening)
\label{thm1}
For any $0<\gamma<1/2-\kappa$, there exists a positive constant $c_1$ such that
\begin{equation}
  \Pr(\max_{1\leqslant j\leqslant p}|\hat{\omega}_j-\omega_j|\geqslant cn^{-\kappa})\leqslant O(p\exp(-c_1n^{1-2(\kappa+\gamma)})).
  \label{eqnew1}
\end{equation}
Under Condition 1(a), we have
\begin{equation}
  \Pr(\mathcal{D}\subseteq\hat{\mathcal{D}})\geqslant 1-O(s_n\exp(-c_1n^{1-2(\kappa+\gamma)})),
    \label{eqnew2}
\end{equation}
where $s_n$ stands for the cardinality of $\mathcal{D}$.
\end{theorem}
Per the statements of this theorem, we conclude that MrDc-SIS is capable to handle the NP-dimensionality of order, $\log(p) = o(n^{1-2\kappa})$, with the minimal assumptions compared to many other sure independence screening methods. For example, \citet{Li} required that each $\Xbf$ and $\Ybf$ satisfies the sub-exponential tail bound. We don't need to have this assumption because $R^\Xbf(\Xbf)$ and $R^\Ybf(\Ybf)$ automatically fall in $[0,1]^d$, which is a bounded set. The bounded random variables automatically satisfy the sub-Gaussian tail bound and hence it is already stronger than the sub-exponential tail bound.

\begin{theorem}(Rank consistency)
\label{thm2}
Under Condition 1(b), we have
\begin{equation}
\label{eqnew3}
\Pr(\min_{j\in\mathcal{D}}\hat\omega_j - \max_{j \in \mathcal{I}}\hat\omega_j>0) > 1 - O(p\exp(-c_1n^{1-2(\kappa_2+\gamma)})),
\end{equation}
where $0<\gamma<1/2 - \kappa$.
\end{theorem}
The rank consistency is a stronger result than the sure screening property. It shows that when the signal strength gap between active features and inactive features satisfies Condition 1(b), the active features will always rank higher than the inactive features, which ensures us to find a threshold and separate the active and inactive sets with high probability.

\subsection{The choice of a threshold}

After ranking the predictors from the most important to the least important utilizing $\hat{\omega}_j$, a threshold is needed so that the active set can be separated from the inactive set, and in turn the set $\hat{\mathcal{D}}$ can be selected. Several studies have been made to suggest a threshold, and they fall into two categories.

The first category is a hard threshold \citep{sis, Li, liu2014feature}. This method suggests to select $d$ variables with the largest dependence score $\hat\omega$, where $d$ is often a multiplier of $[n/\log(n)]$ or $[n^{4/5}/\log(n^{4/5})]$, in which $[a ]$ stands for the integer part of $a$.

The other category is a soft threshold \citep{pan2018generic, liu2020model}. We can generate some auxiliary variables that are completely independent from the response, and calculate the dependence score between the response and these auxiliary variables, then we set the threshold to be the largest dependence score among these ``fake'' dependence scores. Recently, \citet{liu2020model} suggested to use knockoff features, and further select the threshold that can control the false discovery rate (FDR) to a given level. 

We adopt another soft threshold, the max-ratio criterion that was proposed by \citet{huang2014feature} in the MrDc-SIS approach. If sorting the scores $\hat{\omega}_j$'s in descending order, denote as $\hat{\omega}_{(1)}\geqslant \hat{\omega}_{(2)}\geqslant\cdots\geqslant \hat{\omega}_{(p)}$, the importance of predictors can be ranked. \citet{huang2014feature} assumed that $\hat{\omega}_{(j)}>0$ for $j\leqslant s_0$ and $\hat{\omega}_{(j)}\to 0$ in probability for $j>s_0$, here the true selection size is $|\mathcal{D}| = s_0$. It implies that the ratio $\hat{\omega}_{(s_0)}/\hat{\omega}_{(s_0+1)}\to \infty$ in probability. Therefore, $s_0$ can be estimated by
\begin{equation*}
  \hat{s}_0 = \argmax_{1\leqslant j\leqslant p-1}~\hat{\omega}_{(j)}/\hat{\omega}_{(j+1)}.
\end{equation*}

To save the computational cost, we utilize different multipliers of $[n/\log(n)]$ criterion for all simulation studies and the max-ratio rule for the real data analysis.

\section{Simulation studies}

In this section, we evaluate the finite sample performance of MrDc-SIS through three simulation examples, and compare it with eight most relevant feature screening methods. All the approaches are implemented using Python 3, and  we set the hyperparameter $a = 0.5$ for SC-SIS. Similar to \citet{Li} and \citet{dcisis}, we adopt the following three ``efficiency'' criteria to assess the performance of each feature screening method:

\begin{enumerate}
\item $\mathcal{S}$: the minimum selection size to include all true predictors. We draw boxplots of $\mathcal{S}$ across 200 replicates, and also report the mean and standard deviation of $\mathcal{S}$. The closer of $\mathcal{S}$ to the true model size, the robuster the procedure is.
\item $\mathcal{P}_s$: the success rate that each true predictor is selected under three given thresholds across 200 replicates. Following \citet{sis} and \citet{Li}, the thresholds are set as $d_1 = [n/\log(n)]$, $d_2 = 2\times d_1$ and $d_3 = 3\times d_1$.
\item $\mathcal{P}_a$: the simultaneous success rate that all true predictors are selected under three given thresholds across 200 replicates. A higher $\mathcal{P}_s$ (or $\mathcal{P}_a$) value indicates that the procedure has a higher chance to include each true individual predictor (or all true predictors) within a given threshold.
\end{enumerate}

Before conducting simulations, we first assess the computational costs of these nine approaches. We utilize the Hungarian algorithm to solve the map transportation problem, which has the same computational cost, $O(n^3)$, as the PC-Screen does, only under its worse cases. On the contrary,  SC-SIS and DC-SIS have computational costs of $O(n^2)$. We run the ``timeit'' function in a laptop (\emph{Intel Core i7-4720HQ, 16GB RAM, Windows 10}) to compare the actual computational times of each of these nine approaches using 200 observations and 100 iterations for one predictor as an example. In univariate scenarios, both $X$ and $Y$ are $200\times 1$, and in multivariate scenarios, both $\Xv$ and $\Yv$ are $200\times 3$.

\begin{table}[H]
        \centering
        \caption{Computational time cost to calculate one dependence score for each of the nine feature screening approaches. SIS, SIRS, RRCS and DC-RoSIS are not applicable for multivariate scenarios.}
        \label{tbltime}
        \begin{tabular}{ccccccccccc}
        \toprule
             & & \multirow{2}{*}{SIS} & \multirow{2}{*}{SIRS} & \multirow{2}{*}{RRCS} & DC- & DC- & BCor- & PC- & SC- & MrDc- \\
             & & & & & SIS & RoSIS & SIS & Screen & SIS & SIS \\
             \cmidrule(lr){3-3} \cmidrule(lr){4-4} \cmidrule(lr){5-5} \cmidrule(lr){6-6} \cmidrule(lr){7-7} \cmidrule(lr){8-8} \cmidrule(lr){9-9} \cmidrule(lr){10-10} \cmidrule(lr){11-11}
           Univariate   & Mean (ms) & 0.161 & 5.74 & 0.501 & 0.908 & 8.97 & 3.65 & 555 & 5.57 & 17 \\
          ($n = 200$) & Std (ms) & 0.006 & 0.165 & 0.003 & 0.003 & 0.039 & 0.015 & 0.86 & 0.010 & 0.051 \\

           Multivariate  & Mean (ms) & N.A. & N.A. & N.A. & 1.36 & N.A. & 65.5 & 1,370 & 14.3 & 22.1 \\
          ($n = 200$) & Std (ms) & N.A. & N.A. & N.A. & 0.005 & N.A. & 0.321 & 4.27 & 0.017 & 0.234 \\
           \bottomrule
        \end{tabular}
    \end{table}
From Table \ref{tbltime}, we observe that SIS, RRCS and DC-SIS are extremely fast for the univariate case, which require less than 1 ms to calculate a score; BCor-SIS, SC-SIS, SIRS, DC-RoSIS are a little bit slower, requires less than 10 ms to get a score; MrDc-SIS needs 17 ms, and PC-Screen spends 555 ms. In the multivariate case, DC-SIS is the fastest with its speed less than 2 ms; SC-SIS, MrDc-SIS and BCor-SIS are similar; but PC-Screen spends 1370 ms to conduct one score, which is much slower than other four methods.

\subsection{Univariate case}
\begin{example}
Consider the model
\begin{equation*}
 Y =   \beta_1X_1+ \beta_2X_6+ \beta_3X_{12}^2+\beta_4X_{22}+\epsilon,
\end{equation*}
and we set the sample size $n=200$ and the number of predictors $p = 5000$. We generate $\{X_i\}_{i=1}^p$ from a multivariate t distribution with degree of freedom 1 ($t_1$), mean $\0v$, and covariance matrix $\Sigmav_{p\times p}=(\sigma_{ij})$, where $\sigma_{ij} = 0.5^{|i-j|}$. We generate $\beta_j\sim \mathrm{Uniform} (2, 5)$ for $j = 1,2,3,4$. We vary two cases for the error term $\epsilon$. Case (1): $\epsilon\sim N(0, 1)$ and case (2): $\epsilon \sim t_1$ to explore normal errors and heavy tail errors.
\end{example}

\begin{figure}[H]
\begin{minipage}{.48\textwidth}
\centering
\includegraphics[width=\textwidth]{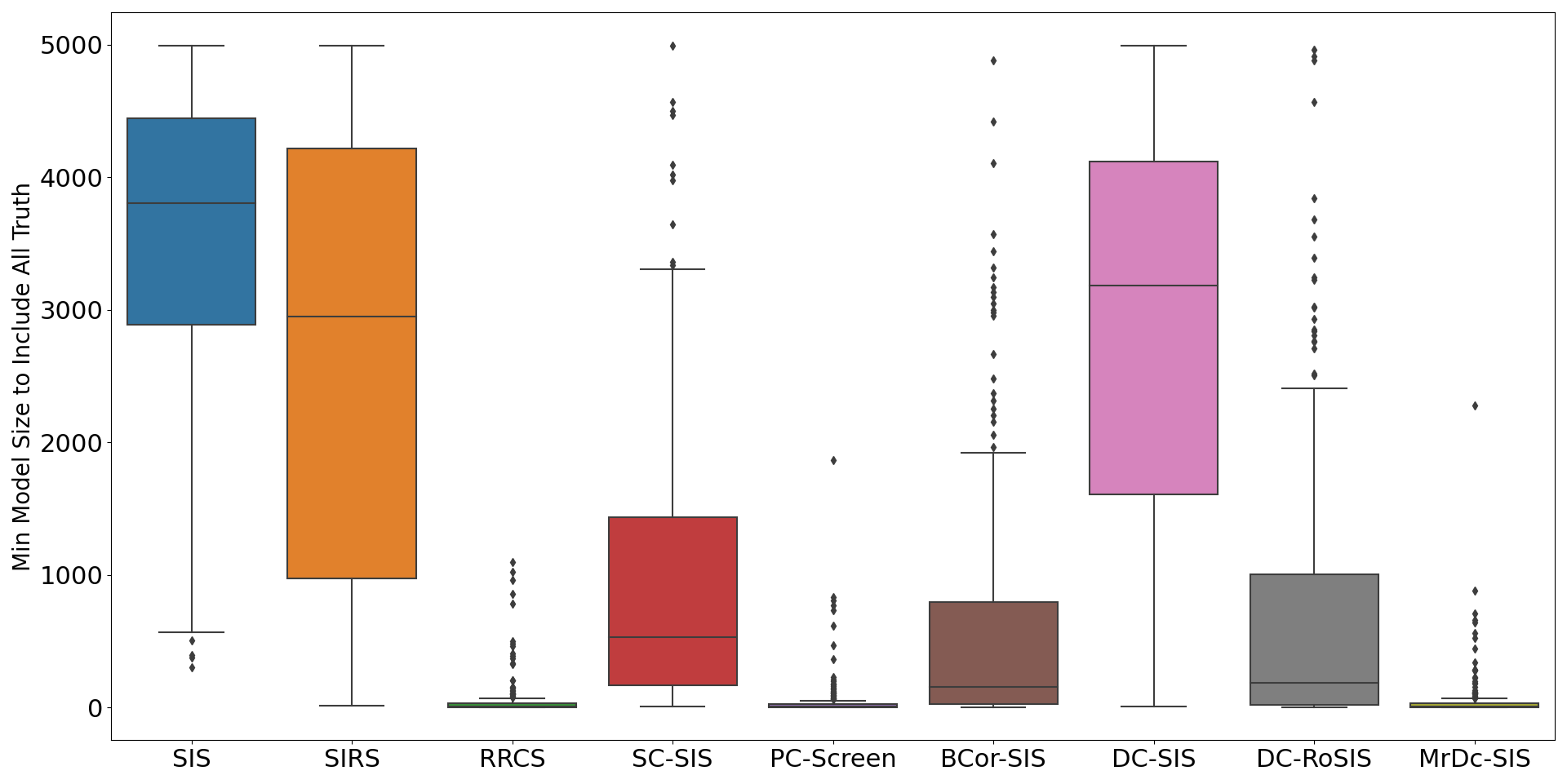}
\caption{Boxplots of $\mathcal{S}$ in case (1) where $\epsilon\sim N(0, 1)$. RRCS, PC-Screen and MrDc-SIS outperform the other methods as the true signals are ranked top among all predictors, while for other methods, like SIS, true signals are ranked around 4000.}
\label{figt1normal}
\end{minipage}
\hspace{1cm}
\begin{minipage}{.48\textwidth}
\centering
\includegraphics[width=\textwidth]{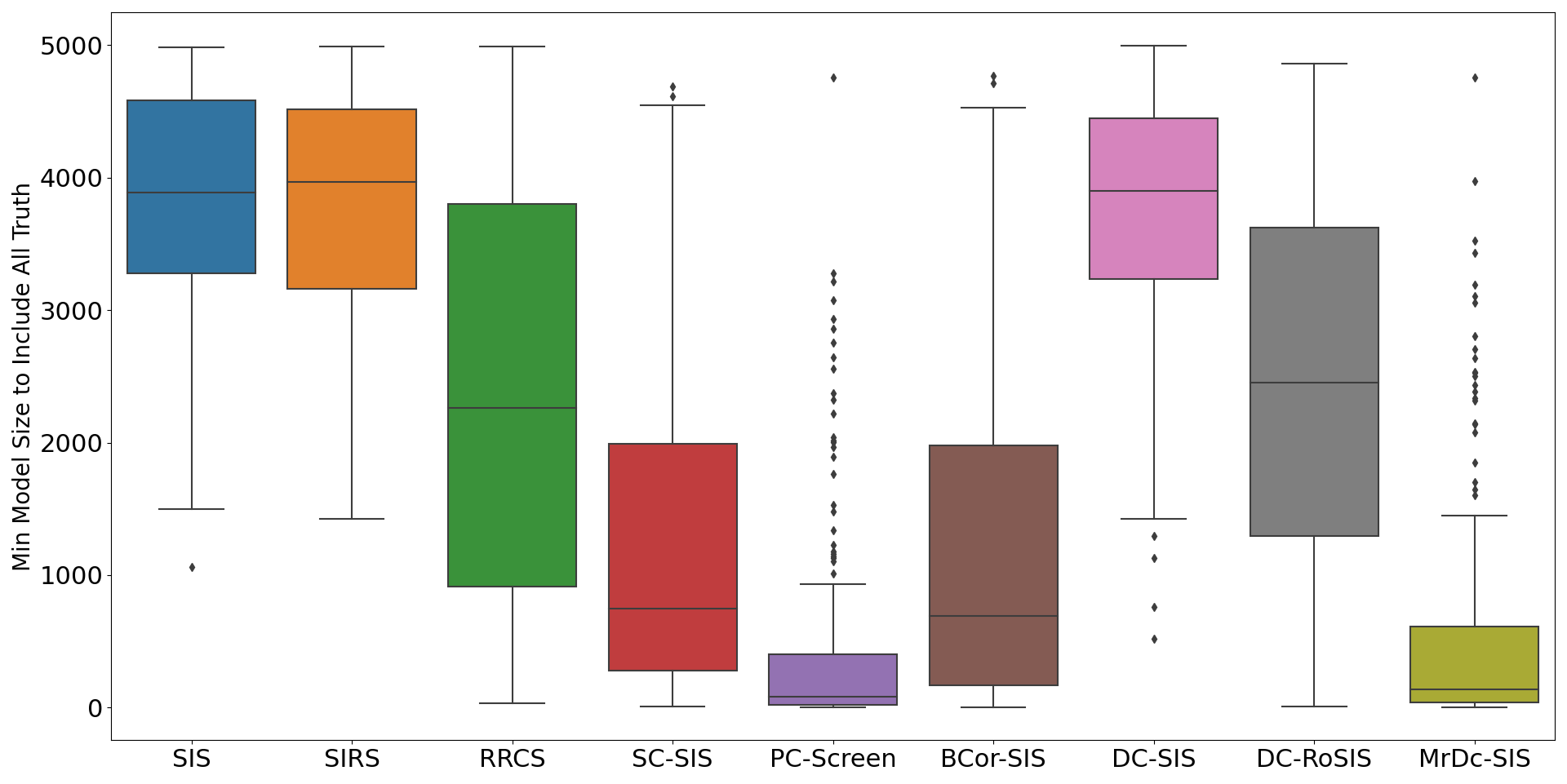}
\caption{Boxplots of $\mathcal{S}$ in case (2) where $\epsilon \sim t_1$. As the noise has heavier tails, all methods get worse than in case (1). However, PC-Screen and MrDc-SIS are still better than other methods, and RRCS performs much worse than before.}
\label{figt1t1}
\end{minipage}
\end{figure}

\begin{table}[H]
    \centering
    \caption{The minimum selection size $\mathcal{S}$ of each screening method for case (1) of Example 1. Among 5000 predictors, RRCS, PC-Screen and MrDc-SIS can rank all true signals to top 60, while SC-SIS, BCor-SIS and DC-RoSIS will rank all true signals to top 1000, and others will rank true signals to 3000, indistinguishable from the noise.}
    \label{tblt1normal1}
    \begin{tabular}{cccccccccc}
    \toprule
           & \multirow{2}{*}{SIS} & \multirow{2}{*}{SIRS} & \multirow{2}{*}{RRCS} & SC- & PC- & BCor- & DC- & DC- & MrDc- \\
              & & & & SIS & Screen & SIS & SIS & RoSIS & SIS \\
        \cmidrule(lr){2-2} \cmidrule(lr){3-3} \cmidrule(lr){4-4} \cmidrule(lr){5-5} \cmidrule(lr){6-6} \cmidrule(lr){7-7} \cmidrule(lr){8-8} \cmidrule(lr){9-9} \cmidrule(lr){10-10} 
       $\mathcal{S}$.mean  & 3494.35  & 2661.64  & \textbf{}{61.31}  & 1004.48  & \textbf{}{56.81}  & 643.93  & 2856.78  & 766.64  & \textbf{}{62.31}  \\
       $\mathcal{S}$.std & 1221.44 & 1662.03 & 165.64 & 1108.24 & 181.27 & 988.96 & 1469.41 & 1118.37 & 201.52 \\
       \bottomrule
    \end{tabular}
\end{table}
\begin{table}[H]
\centering
\caption{$\mathcal{P}_s$ and $\mathcal{P}_a$ under different thresholds for case (1) of Example 1, $d_1 = [n/\log(n)] = 37$, $d_2 = 2d_1$, and $d_3 = 3d_1$. If we choose top $d_3$ predictors, RRCS, PC-Screen and MrDc-SIS can select all true signals 90\% of the time, BCor-SIS and DC-RoSIS can select all true signals 45\% of the time, while approaches that are not robust, like SIS and SIRS, only 6\% of the time they can select all true signals.}
\label{tblt1normal2}
\scriptsize

\begin{tabular}{*{16}{c}}
  \toprule
  \multirow{2}{*}{Model Size} &  \multicolumn{5}{c}{SIS}  & \multicolumn{5}{c}{SIRS} & \multicolumn{5}{c}{RRCS} \\ 
  \cmidrule(lr){2-6} \cmidrule(lr){7-11} \cmidrule(lr){12-16}
   &   \multicolumn{4}{c} {$\mathcal{P}_s$} & $\mathcal{P}_a$ & \multicolumn{4}{c} {$\mathcal{P}_s$} & $\mathcal{P}_a$ & \multicolumn{4}{c} {$\mathcal{P}_s$} & $\mathcal{P}_a$ \\
   \cmidrule(lr){2-5} \cmidrule(lr){6-6} \cmidrule(lr){7-10} \cmidrule(lr){11-11} \cmidrule(lr){12-15} \cmidrule(lr){16-16}
  & $X_1$ & $X_6$ & $X_{12}$ & $X_{22}$ & All & $X_1$ & $X_6$ & $X_{12}$ & $X_{22}$ & All & $X_1$ & $X_6$ & $X_{12}$ & $X_{22}$ & All  \\
  $d_1$ & 0.125 & 0.16 & 0.1 & 0.11 & 0 & 0.335 & 0.325 & 0.32 & 0.275 & 0.025  & 0.955 & 0.96 & 0.93 & 0.92 & 0.775  \\
  $d_2$ & 0.15 & 0.195 & 0.125 & 0.155 & 0  & 0.415 & 0.395 & 0.415 & 0.365 & 0.045 & 0.97 & 0.975 & 0.96 & 0.95 & 0.86  \\
  $d_3$ & 0.17 & 0.235 & 0.165 & 0.155 & 0 & 0.445 & 0.435 & 0.45 & 0.39 & 0.06 & 0.98 & 0.975 & 0.97 & 0.975 & \textbf{0.9}   \\
  \midrule
  \multirow{2}{*}{Model Size} & \multicolumn{5}{c}{SC-SIS} & \multicolumn{5}{c}{PC-Screen} &  \multicolumn{5}{c}{BCor-SIS} \\
  \cmidrule(lr){2-6} \cmidrule(lr){7-11} \cmidrule(lr){12-16}
   &   \multicolumn{4}{c} {$\mathcal{P}_s$} & $\mathcal{P}_a$ & \multicolumn{4}{c} {$\mathcal{P}_s$} & $\mathcal{P}_a$ & \multicolumn{4}{c} {$\mathcal{P}_s$} & $\mathcal{P}_a$ \\
  \cmidrule(lr){2-5} \cmidrule(lr){6-6} \cmidrule(lr){7-10} \cmidrule(lr){11-11} \cmidrule(lr){12-15} \cmidrule(lr){16-16}
  & $X_1$ & $X_6$ & $X_{12}$ & $X_{22}$ & All & $X_1$ & $X_6$ & $X_{12}$ & $X_{22}$ & All & $X_1$ & $X_6$ & $X_{12}$ & $X_{22}$ & All  \\
  $d_1$ & 0.65 & 0.645 & 0.605 & 0.56 & 0.08  & 0.965 & 0.96 & 0.955 & 0.91 & 0.8 & 0.77 & 0.76 & 0.79 & 0.735 & 0.27 \\
  $d_2$ & 0.7 & 0.69 & 0.655 & 0.615 & 0.125 & 0.965 & 0.97 & 0.955 & 0.955 & 0.85 & 0.82 & 0.8  & 0.81 & 0.785 & 0.365\\
  $d_3$ & 0.745 & 0.71 & 0.7 & 0.66 & 0.175 & 0.975 & 0.98 & 0.975 & 0.965 & \textbf{0.895} & 0.86 & 0.855 & 0.825 & 0.805 & 0.455 \\
  \midrule
  \multirow{2}{*}{Model Size} & \multicolumn{5}{c}{DC-SIS} & \multicolumn{5}{c}{DC-RoSIS} & \multicolumn{5}{c}{MrDc-SIS} \\
  \cmidrule(lr){2-6} \cmidrule(lr){7-11} \cmidrule(lr){12-16}
   & \multicolumn{4}{c} {$\mathcal{P}_s$} & $\mathcal{P}_a$ & \multicolumn{4}{c} {$\mathcal{P}_s$} & $\mathcal{P}_a$ & \multicolumn{4}{c} {$\mathcal{P}_s$} & $\mathcal{P}_a$ \\
  \cmidrule(lr){2-5} \cmidrule(lr){6-6} \cmidrule(lr){7-10} \cmidrule(lr){11-11} \cmidrule(lr){12-15} \cmidrule(lr){16-16}
    & $X_1$ & $X_6$ & $X_{12}$ & $X_{22}$ & All & $X_1$ & $X_6$ & $X_{12}$ & $X_{22}$ & All & $X_1$ & $X_6$ & $X_{12}$ & $X_{22}$ & All  \\
  $d_1$  & 0.21 & 0.235 & 0.195 & 0.18 & 0.01 & 0.735 & 0.715 & 0.705 & 0.65 & 0.32 & 0.965 & 0.96  & 0.935 & 0.905 & 0.775 \\
  $d_2$  & 0.265 & 0.28 & 0.255 & 0.22 & 0.01 & 0.77 & 0.76 & 0.755 & 0.72 & 0.4 & 0.97 & 0.965 & 0.955 & 0.95 & 0.845 \\
  $d_3$  & 0.305 & 0.33 & 0.28 & 0.25 & 0.015 & 0.795 & 0.81 & 0.775 & 0.74 & 0.44 & 0.975 & 0.975 & 0.97 & 0.97 & \textbf{0.89} \\
  \bottomrule

\end{tabular}
\end{table}

Tables \ref{tblt1normal1}, \ref{tblt1normal2} and Fig.~\ref{figt1normal} demonstrate the results of all the nine approaches in Example 1 where the error term $\epsilon\sim N(0,1)$. It can be observed from Fig.~\ref{figt1normal} that RRCS, PC-Screen and MrDc-SIS outperform the other screening approaches, and the minimum selection size for these three methods to select all true predictors is around 60. Given the fact that the total number of predictors is 5000, average $\mathcal{S}$'s bigger than 2500 for SIS, SIRS and DC-SIS imply that they can not distinguish the truth from the noise; SC-SIS, BCor-SIS and DC-RoSIS are all robust screening methods, and they perform better than SIS, SIRS and DC-SIS, however, their average $\mathcal{S}$'s are still ten times higher than MrDc-SIS; We can also observe from Table~\ref{tblt1normal2} that the simultaneous success rates $\mathcal{P}_a$'s of SIS, SIRS, and DC-SIS are just a little above zero even for a threshold of $d_3$, the $\mathcal{P}_a$'s of SC-SIS, Bcor-SIS, and DC-RoSIS are less than 0.45 for a threshold $d_3$, while RRCS, PC-Screen, and MrDc-SIS achieve success rates of more than 0.89 under the same threshold. The results in case (1) indicate that the MrDc-SIS is as robust as RRCS and PC-Screen do, and outperforms the other six feature screening methods when error has the normal distribution.

\begin{table}[H]
    \centering
    \caption{The minimum selection size $\mathcal{S}$ of each screening method for case (2) of Example 1. All methods perform worse than in case (1), but PC-Screen and MrDc-SIS still outperform the others, which can rank all true signals at around 500.}
    \label{tblt1t11}
    \begin{tabular}{cccccccccc}
    \toprule
           & \multirow{2}{*}{SIS} & \multirow{2}{*}{SIRS} & \multirow{2}{*}{RRCS} & SC- & PC- & BCor- & DC- & DC- & MrDc- \\
              & & & & SIS & Screen & SIS & SIS & RoSIS & SIS \\
         \cmidrule(lr){2-2} \cmidrule(lr){3-3} \cmidrule(lr){4-4} \cmidrule(lr){5-5} \cmidrule(lr){6-6} \cmidrule(lr){7-7} \cmidrule(lr){8-8} \cmidrule(lr){9-9} \cmidrule(lr){10-10} 
       $\mathcal{S}$.mean  & 3807.90  & 3793.81  & 2378.06  & 1297.39  & \textbf{}{421.63}  & 1187.96  & 3720.05  & 2441.23  & \textbf{}{538.07}  \\
       $\mathcal{S}$.std & 863.91 & 895.86 & 1522.62 & 1239.26 & 776.74 & 1227.52 & 945.82 & 1338.28 & 872.06 \\
       \bottomrule
    \end{tabular}
\end{table}

\begin{table}[H]
\centering
\caption{$\mathcal{P}_s$ and $\mathcal{P}_a$ under different thresholds for case (2) of Example 1, $d_1 = [n/\log(n)] = 37$, $d_2 = 2d_1$, and $d_3 = 3d_1$. If we choose top $d_3$ predictors, PC-Screen and MrDc-SIS can select all true signals in 50\% of the time, SC-SIS and BCor-SIS can select all true signals in 10\% of the time, and others can only successfully select all true signals in less than 2\% of the time.}
\label{tblt1t12}
\scriptsize
\begin{tabular}{*{16}{c}}
  \toprule
  \multirow{2}{*}{Model Size} &  \multicolumn{5}{c}{SIS}  & \multicolumn{5}{c}{SIRS} & \multicolumn{5}{c}{RRCS}   \\
  \cmidrule(lr){2-6} \cmidrule(lr){7-11} \cmidrule(lr){12-16}
   &   \multicolumn{4}{c} {$\mathcal{P}_s$} & $\mathcal{P}_a$ & \multicolumn{4}{c} {$\mathcal{P}_s$} & $\mathcal{P}_a$ & \multicolumn{4}{c} {$\mathcal{P}_s$} & $\mathcal{P}_a$  \\
   \cmidrule(lr){2-5} \cmidrule(lr){6-6} \cmidrule(lr){7-10} \cmidrule(lr){11-11} \cmidrule(lr){12-15} \cmidrule(lr){16-16}
  & $X_1$ & $X_6$ & $X_{12}$ & $X_{22}$ & All & $X_1$ & $X_6$ & $X_{12}$ & $X_{22}$ & All & $X_1$ & $X_6$ & $X_{12}$ & $X_{22}$ & All  \\
  $d_1$ & 0 & 0.005 & 0.095 & 0.005 & 0 & 0.05 & 0.035 & 0.015 & 0.035 & 0  & 0.65 & 0.645 & 0.01 & 0.73 & 0.005  \\
  $d_2$ & 0.005 & 0.015 & 0.13 & 0.02 &  0 & 0.06 & 0.05 & 0.03 & 0.045 & 0 & 0.73 & 0.74 & 0.03 & 0.765 & 0.02  \\
  $d_3$ & 0.005 & 0.015 & 0.175 & 0.03 & 0 & 0.06 & 0.085 & 0.045 & 0.065 & 0 & 0.76 & 0.77 & 0.04 & 0.79 & 0.025   \\
  \midrule
\multirow{2}{*}{Model Size} & \multicolumn{5}{c}{SC-SIS} & \multicolumn{5}{c}{PC-Screen} &  \multicolumn{5}{c}{BCor-SIS} \\
\cmidrule(lr){2-6} \cmidrule(lr){7-11} \cmidrule(lr){12-16}
& \multicolumn{4}{c} {$\mathcal{P}_s$} & $\mathcal{P}_a$ & \multicolumn{4}{c} {$\mathcal{P}_s$} & $\mathcal{P}_a$ & \multicolumn{4}{c} {$\mathcal{P}_s$} & $\mathcal{P}_a$ \\
\cmidrule(lr){2-5} \cmidrule(lr){6-6} \cmidrule(lr){7-10} \cmidrule(lr){11-11} \cmidrule(lr){12-15} \cmidrule(lr){16-16}
& $X_1$ & $X_6$ & $X_{12}$ & $X_{22}$ & All & $X_1$ & $X_6$ & $X_{12}$ & $X_{22}$ & All & $X_1$ & $X_6$ & $X_{12}$ & $X_{22}$ & All \\
$d_1$ & 0.48 & 0.4  & 0.835 & 0.455 & 0.03  & 0.755 & 0.745 & 0.97 & 0.8 & 0.365 & 0.535 & 0.495 & 1 & 0.52 & 0.065 \\
$d_2$ & 0.57 & 0.5 & 0.91 & 0.53 & 0.08 & 0.795 & 0.82 & 0.985 & 0.825 & 0.475 & 0.58 & 0.545 & 1 & 0.59 & 0.125 \\
$d_3$ & 0.61 & 0.56 & 0.925 & 0.545 & 0.1 & 0.815 & 0.87 & 0.995 & 0.86 & \textbf{0.565} & 0.625 & 0.6 & 1 & 0.635 & 0.175 \\
\midrule
  \multirow{2}{*}{Model Size}   & \multicolumn{5}{c}{DC-SIS} & \multicolumn{5}{c}{DC-RoSIS} & \multicolumn{5}{c}{MrDc-SIS}  \\
  \cmidrule(lr){2-6} \cmidrule(lr){7-11} \cmidrule(lr){12-16}
   &   \multicolumn{4}{c} {$\mathcal{P}_s$} & $\mathcal{P}_a$ & \multicolumn{4}{c} {$\mathcal{P}_s$} & $\mathcal{P}_a$ & \multicolumn{4}{c} {$\mathcal{P}_s$} & $\mathcal{P}_a$ \\
   \cmidrule(lr){2-5} \cmidrule(lr){6-6} \cmidrule(lr){7-10} \cmidrule(lr){11-11} \cmidrule(lr){12-15} \cmidrule(lr){16-16}
    & $X_1$ & $X_6$ & $X_{12}$ & $X_{22}$ & All & $X_1$ & $X_6$ & $X_{12}$ & $X_{22}$ & All & $X_1$ & $X_6$ & $X_{12}$ & $X_{22}$ & All   \\
  $d_1$  & 0.005 & 0 & 0.17 & 0.01 & 0 & 0.205 & 0.175 & 0.585 & 0.2 & 0.005 & 0.7  & 0.665 & 0.93 & 0.73 & 0.22 \\
  $d_2$  & 0.005 & 0.02 & 0.24 & 0.015 & 0 & 0.265 & 0.215 & 0.635 & 0.225 & 0.01 & 0.76 & 0.745 & 0.98 & 0.8 & 0.385\\
  $d_3$  & 0.025 & 0.03 & 0.28 & 0.02 & 0 & 0.31 & 0.24 & 0.66 & 0.255 & 0.02 & 0.78 & 0.8 & 0.985 & 0.825 & \textbf{0.47} \\
  \bottomrule
  
\end{tabular}

\end{table}

Tables \ref{tblt1t11}, \ref{tblt1t12} and Fig.~\ref{figt1t1} demonstrate the results of all the nine approaches in Example 1 where the error term $\epsilon\sim t_1$. This time not only the predictors are messier than the normal case, but also the error terms have heavier tails than the normal distribution. As a result, all methods perform worse than in case (1). Among these nine approaches, PC-Screen performs the best; MrDc-SIS is the second with an average $\mathcal{S}$ of 538; SC-SIS and Bcor-SIS are in the next tier, with average $\mathcal{S}$'s being around 1200; RRCS and DC-RoSIS have average $\mathcal{S}$'s around 2400; and SIS, SIRS and DC-SIS are in the last tier with average $\mathcal{S}$'s around 3700. We can also observe from Table~\ref{tblt1t12} that the simultaneous success rates, $\mathcal{P}_a$'s, of SIS, SIRS, and DC-SIS are 0 even in threshold $d_3$, and the $\mathcal{P}_s$'s of any individual true predictor are small; DC-RoSIS performs better than DC-SIS in individual $\mathcal{P}_s$'s, but its simultaneous $\mathcal{P}_a$ is only 0.02 at the threshold $d_3$; RRCS has good performance on individuals like $X_1, X_6$ and $X_{22}$, however, it can not handle the square term $X_{12}^2$ since it may not be monotone, and its simultaneous rate $\mathcal{P}_a$ is only 0.025 at the threshold $d_3$; SC-SIS and BCor-SIS work better as they can successfully select $X_{12}$, and their simultaneous $\mathcal{P}_a$'s increase to 0.1; PC-Screen and MrDc-SIS select each true predictor with a high probability, and the $\mathcal{P}_a$'s of these two methods increase to 0.5. The results in case (2) indicate that the MrDc-SIS is as robust as PC-Screen and outperforms the other seven feature screening methods when both the predictors and the error term have heavy tails like $t_1$ distribution. 

\subsection{Multivariate case}
In this section we evaluate the performance of MrDc-SIS on multivariate scenarios, especially when both responses and predictors are multivariate. Among the nine screening methods, only SC-SIS, PC-Screen, BCor-SIS, DC-SIS and MrDc-SIS can handle multivariate data, therefore, we compare only these five feature screening methods in the following two examples.

\begin{example}(Multivariate $t$ scenario)
We set sample size $n = 200$, predictor size $p = 10,000$, and response size $q= 10$. We generate $\Uv = [U_1,\dots, U_p]$ from multivariate $t_2$ distribution with covariance $\Sigmav_{p\times p}=(\sigma_{ij})$, where $\sigma_{ij} = 0.8^{|i-j|}$, $\Vv = [V_1,\dots, V_p]$ from multivariate $t_1$ distribution with same covariance $\Sigmav_{p\times p}$, and $\Wv = [W_1,\dots,W_p]$ from multivariate $t_3$ distribution with same covariance $\Sigmav_{p\times p}$. Then we integrate them as our predictors $\Xv=[X_1,\dots, X_p]$ where $X_j = [U_j, V_j, W_j]$ for $j \in\{ 1,\dots, p\}$. Here, $\Xv$ is a $3\times 200\times 10,000$ array (tensor), and each $X_j$ is a multivariate vector with dimension $200 \times 3$. This design mimic the real data settings where $\Uv, \Vv, \Wv$ are data collected from different platforms (omes). In addition, it demonstrates the performance of MrDc-SIS when screening multivariate predictors. In the following, we connect the first 4 response components to be truly associated with some active predictors (different predictor connect different platforms to increase difficulty level). We set the remaining 6 response components as noises.
\begin{itemize}
	\item For $k  \in \{1, \dots, 4\}$,
            \begin{enumerate}
                \item Randomly sample $id_1, id_2, id_3, id_4$ with replacement from $\{1,2,3\}$.
                \item $Y_k[i] = \beta_1 X[id_1,i,2]+\beta_2 X[id_2,i,4] + \beta_3 X[id_3, i, 101] + \beta_4 X[id_4, i, 102]^2+\epsilon[i]$, $\forall i = 1,2,\dots,n$, where $\beta_{1,2,3,4}\sim \mathrm{Uniform}(1,2)$ and $\epsilon\sim t_1$.
            \end{enumerate}
            \item For $k\in \{5, \dots, 10\}$, $Y_k\sim t_1$.
\end{itemize}
\end{example}

\begin{figure}[H]
     \centering
     \includegraphics[width = 0.6\textwidth]{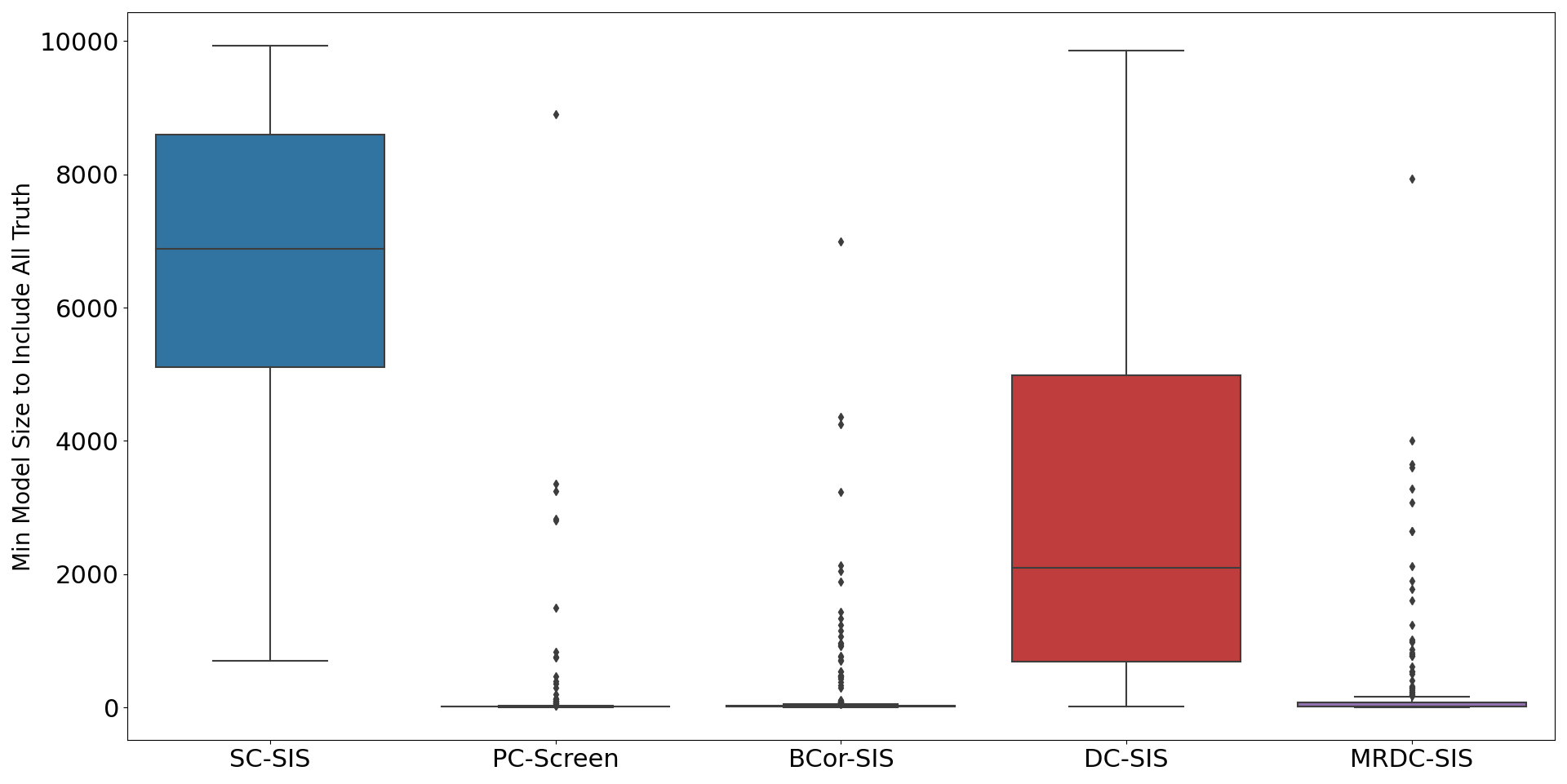}
     \caption{Boxplots of $\mathcal{S}$ in multivariate $t$ scenario. PC-Screen, BCor-SIS and MrDc-SIS perform well under this setting, while SC-SIS and DC-SIS find it hard to distinguish the signals from the noise.}
     \label{figmultit1}
 \end{figure}

\begin{table}[H]
    \centering
    \caption{The minimum selection size $\mathcal{S}$ of each screening method for Example 2. PC-Screen, BCor-SIS and MrDc-SIS can rank all true signals at top 200, while DC-SIS ranks all true signals at top 3000, and SC-SIS ranks all true signals at top 6000.}
    \label{tblmultit1}
    \begin{tabular}{cccccc}
    \toprule
         &  SC-SIS & PC-Screen & Bcor-SIS & DC-SIS & MrDc-SIS \\
         \cmidrule(lr){2-2} \cmidrule(lr){3-3} \cmidrule(lr){4-4} \cmidrule(lr){5-5} \cmidrule(lr){6-6}
       $\mathcal{S}$.mean  & 6603.45  & \textbf{}{146.07}  & 226.23  & 2978.20   & 276.98  \\
       $\mathcal{S}$.std & 2241.11 & 766.61 & 755.20 & 2616.41 & 860.28 \\
       \bottomrule
    \end{tabular}
\end{table}
\begin{table}[H]
\centering
\caption{$\mathcal{P}_s$ and $\mathcal{P}_a$ under different thresholds for Example 2, $d_1 = [n/\log(n)] = 37$, $d_2 = 2d_1$, and $d_3 = 3d_1$. If we choose top $d_3$ predictors, PC-Screen can select all true signals in 92\% of the time, BCor-SIS and MrDc-SIS can select all true signals in 80\% of the time, while SC-SIS and DC-SIS have a much lower success rate.}
\label{tblmultit2}
\scriptsize

\begin{tabular}{*{16}{c}}
  \toprule
  \multirow{2}{*}{Model Size} &  \multicolumn{5}{c}{SC-SIS}  & \multicolumn{5}{c}{PC-Screen} & \multicolumn{5}{c}{BCor-SIS}   \\
  \cmidrule(lr){2-6} \cmidrule(lr){7-11} \cmidrule(lr){12-16}
   &   \multicolumn{4}{c} {$\mathcal{P}_s$} & $\mathcal{P}_a$ & \multicolumn{4}{c} {$\mathcal{P}_s$} & $\mathcal{P}_a$ & \multicolumn{4}{c} {$\mathcal{P}_s$} & $\mathcal{P}_a$  \\
   \cmidrule(lr){2-5} \cmidrule(lr){6-6} \cmidrule(lr){7-10} \cmidrule(lr){11-11} \cmidrule(lr){12-15} \cmidrule(lr){16-16}
  & $X_2$ & $X_4$ & $X_{101}$ & $X_{102}$ & All & $X_2$ & $X_4$ & $X_{101}$ & $X_{102}$ & All & $X_2$ & $X_4$ & $X_{101}$ & $X_{102}$ & All  \\
  $d_1$ & 0.01 & 0.02& 0.02& 0.04 & 0 & 0.945& 0.935& 0.955& 0.96 & 0.87 & 0.89 & 0.835& 0.955& 0.96 & 0.775  \\
  $d_2$ & 0.01 & 0.02 & 0.04 & 0.055 & 0 & 0.965& 0.955& 0.97 & 0.96 & 0.895 & 0.925& 0.865& 0.96 & 0.965 & 0.815  \\
  $d_3$ & 0.01 & 0.03 & 0.045& 0.065 & 0 & 0.98 & 0.955& 0.97 & 0.97 & \textbf{0.915} & 0.935& 0.905& 0.965& 0.965 & 0.85  \\
  \midrule
  \multirow{2}{*}{Model Size} & \multicolumn{5}{c}{DC-SIS} & \multicolumn{5}{c}{MrDc-SIS} & & & & & \\
  \cmidrule(lr){2-6} \cmidrule(lr){7-11} 
  & \multicolumn{4}{c} {$\mathcal{P}_s$} & $\mathcal{P}_a$ & \multicolumn{4}{c} {$\mathcal{P}_s$} & $\mathcal{P}_a$ & & & & & \\
  \cmidrule(lr){2-5} \cmidrule(lr){6-6} \cmidrule(lr){7-10} \cmidrule(lr){11-11} 
  & $X_2$ & $X_4$ & $X_{101}$ & $X_{102}$ & All & $X_2$ & $X_4$ & $X_{101}$ & $X_{102}$ & All & & & & & \\
  $d_1$  & 0.12 & 0.125& 0.75 & 0.805 & 0.03 & 0.98& 0.98& 0.82& 0.75 & 0.68 & & & & & \\
  $d_2$ & 0.165& 0.135& 0.845& 0.855 & 0.055 & 0.98 & 0.985& 0.85 & 0.82 & 0.75 & & & & & \\
  $d_3$ & 0.19 & 0.165& 0.88 & 0.865 & 0.06 & 0.985& 0.985& 0.88 & 0.84 & 0.79 & & & & & \\
  \bottomrule
\end{tabular}
\end{table}

Tables \ref{tblmultit1}, \ref{tblmultit2} and Fig.~\ref{figmultit1} demonstrate the results of all the five approaches in Example 2. We observe that PC-Screen performs the best with an average $\mathcal{S}$ of 146; BCor-SIS and MrDc-SIS are close with average $\mathcal{S}$'s less than 300 (the total number of predictors is 10,000); DC-SIS has an average $\mathcal{S}$ be around 3000; SC-SIS has a $\mathcal{S}$ of 6000. From Table~\ref{tblmultit2} we can see the PC-Screen has a simultaneous success rate $\mathcal{P}_a$ of 0.9 when using threshold $d_3$; BCor-SIS and MrDc-SIS have $\mathcal{P}_a$'s being around 0.8 with the same threshold; and SC-SIS and DC-SIS have $\mathcal{P}_a$'s only a little above 0 with the same threshold. The results in Example 2 indicate that the MrDc-SIS performs slightly worse than PC-Screen and BCor-SIS, but it is still robust to select true predictors under multivariate $t$ scenario.

\begin{example}(Multivariate Pareto scenarios)
In this example, we investigate the performance of MrDc-SIS for not only heavier tails but also right skewness. In addition, we mixed both continuous and discrete data types in both the predictor and response components, which mimic the real multi-omics data better and also greatly expand the application scope of the proposed approach. We set predictor size $p = 5000$ and keep $n = 200$ as well as $q = 10$. For a Pareto distribution $Pareto(a, m)$, the probability density is
\begin{equation*}
f(x) = \frac{am^a}{x^{a+1}}
\end{equation*}
where $a$ is the shape and $m$ the scale.

Case (1) (continuous responses).
\begin{enumerate}
\item Generate $\Uv = [U_1,\dots, U_p]$ from multivariate Pareto distribution with shape $a =10$, scale $m =15$, and covariance matrix $\Sigmav_{p\times p}=(\sigma_{ij})$, where $\sigma_{ij} = 0.8^{|i-j|}$.
\item Generate $\Vv = [V_1, \dots, V_p]$ from binomial distribution $\mathrm{Bin}(4, 0.3)$.
\item Generate $\Wv = [W_1,\dots, W_p]$ from multivariate Pareto distribution with shape $a =12$, scale $m =30$, and covariance matrix $\Sigmav_{p\times p}=(\sigma_{ij})$, where $\sigma_{ij} = 0.8^{|i-j|}$.
\item $\Xv=[X_1,\dots, X_p]$ where $X_j = [U_j, V_j, W_j]$ for $ j \in\{ 1,\dots, p\}$. Here, $\Xv$ is a $3\times 200\times 10000$ array (tensor).
\item Connect the first four response components with active predictors as follows
\begin{itemize}
\item For $k\in \{1, \dots, 4\}$,
\begin{enumerate}
\item Randomly sample $id_1, id_2, id_3, id_4$ with replacement from $\{1,2,3\}$.
\item $Y_k[i] = \beta_1 X[id_1,i,2]+\beta_2 X[id_2,i,3] + \beta_3 X[id_3, i, 101] + \beta_4 X[id_4, i, 102]^2+\epsilon[i]$, $ i \in\{ 1,\dots,n\}$, where $\beta_{1,2,3,4}\sim \mathrm{Uniform}(1,2)$ and $\epsilon\sim \mathrm{Pareto}(10,15)$.
\end{enumerate}
\item For $k \in\{ 5, \dots, 10\}$, $Y_k\sim \mathrm{Pareto}(10,15)$.
\end{itemize}
\end{enumerate}

Case (2) (mixed categorical and continuous responses).
We follow exactly the same procedure as in case (1) in generating $\Xv$ and $\Yv$. But we discretize the second component of $\Yv$ as $Y_2[i]=0$ if its value is below 25\% quantile; $Y_2[i]=1$ if its value is between 25\% and 50\% quantile; $Y_2[i]=2$ if its value is between 50\% and 75\% quantile; and $Y_2[i]=3$ if its value is above 75\% quantile. 

\end{example}

\begin{figure}[H]
\begin{minipage}{.48\textwidth}
\centering
\includegraphics[width=\textwidth]{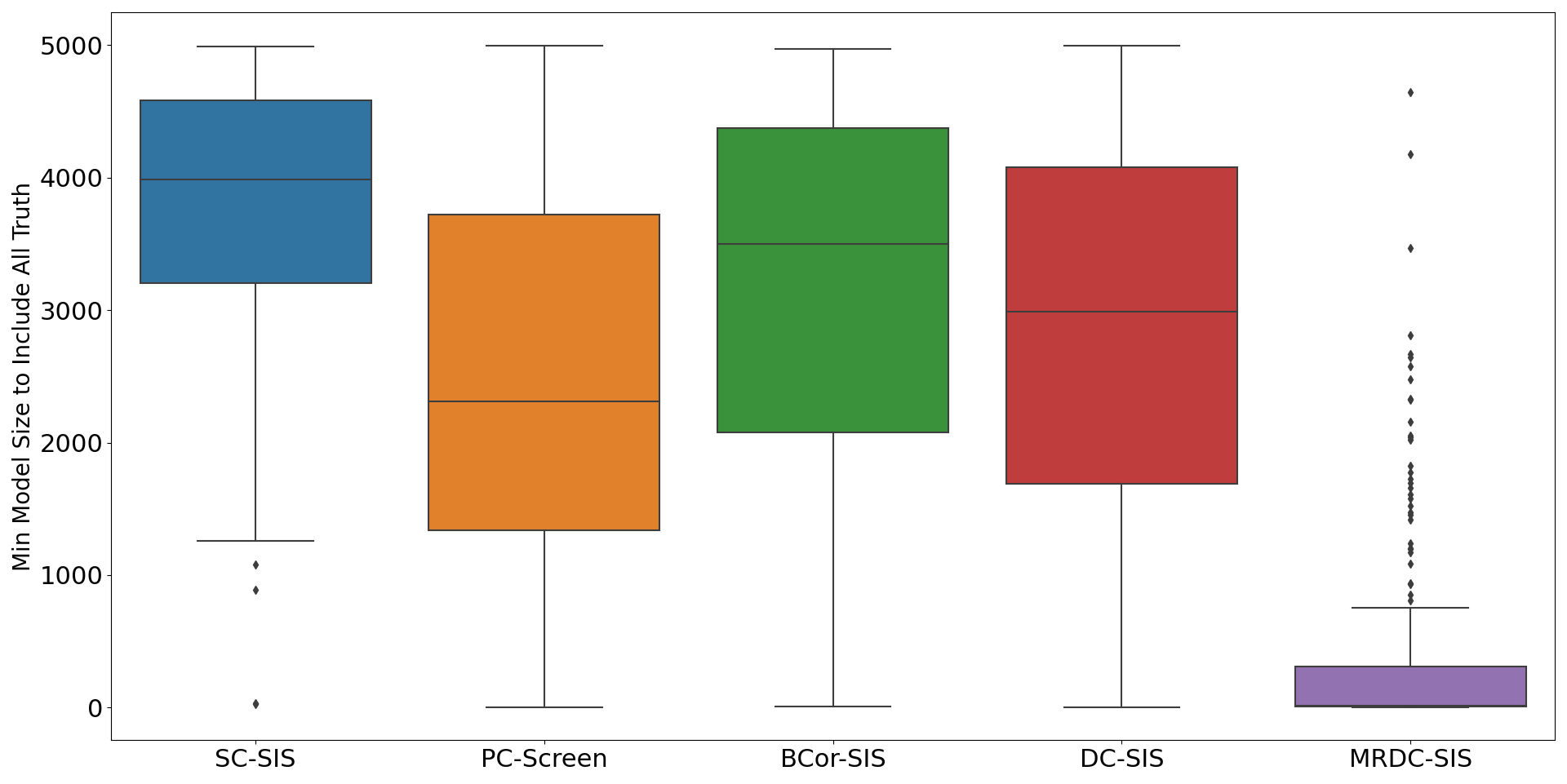}
\caption{Boxplots of $\mathcal{S}$ in case (1) of Example 3. MrDc-SIS performs the best among all five methods, it can rank all true signals in top 400, while other four methods rank all true signals to 4000, implies they have difficulty in distinguishing the signals from the noise.}
\label{figpareto1}
\end{minipage}
\hspace{1cm}
\begin{minipage}{.48\textwidth}
\centering
\includegraphics[width=\textwidth]{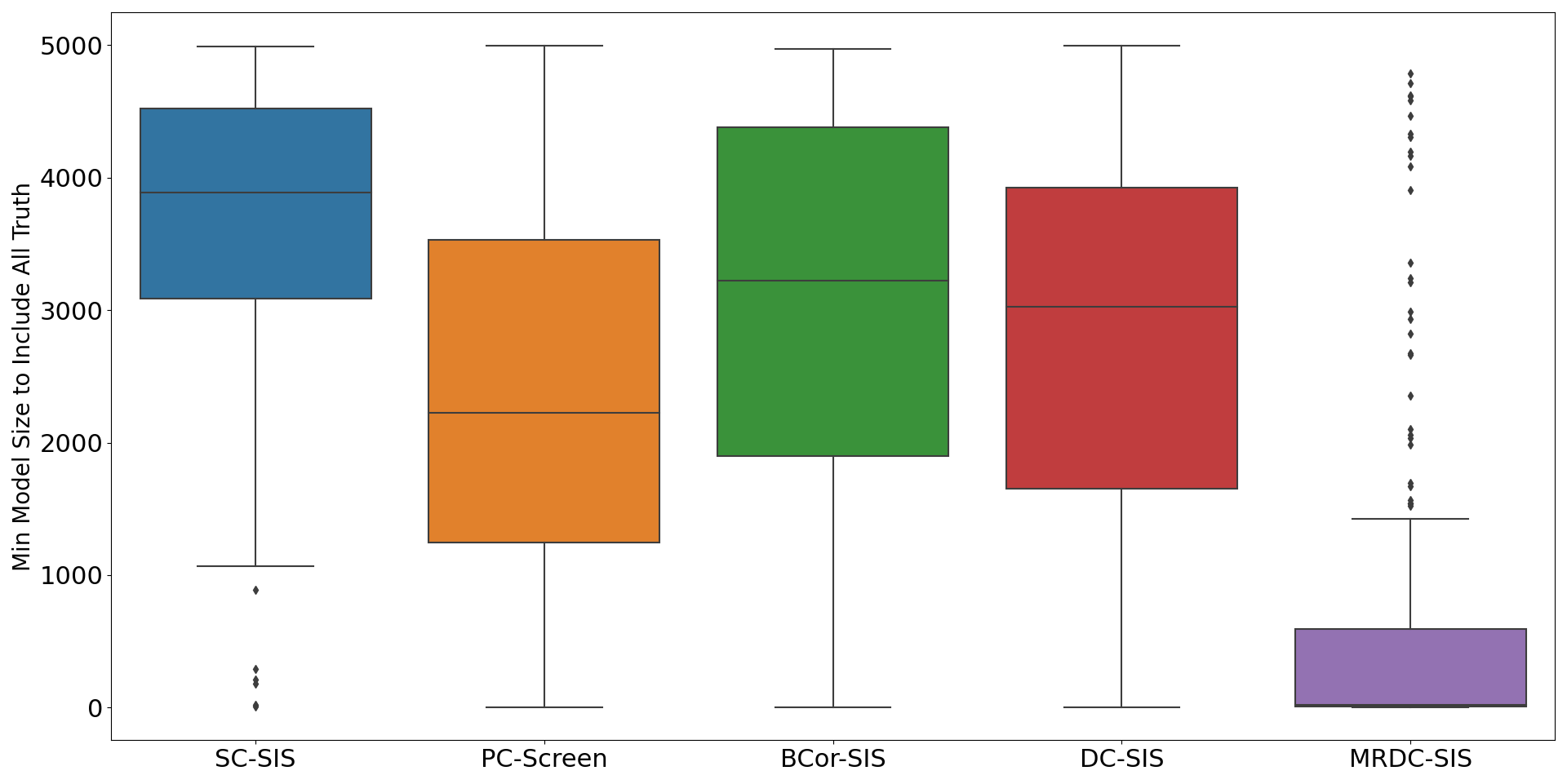}
\caption{Boxplots of $\mathcal{S}$ in case (2) of Example 3. MrDc-SIS still performs the best, it can rank all true signals in top 600, while other four methods again rank all true signals to 4000, implies they have difficulty in distinguishing the signals from the noise.}
\label{figpareto2}
\end{minipage}
\end{figure}

 \begin{table}[H]
    \centering
    \caption{The minimum selection size $\mathcal{S}$ of each screening method for case (1) of Example 3. MrDc-SIS ranks true signals in top 400, better than other four approaches.}
    \label{tblpareto11}
    \begin{tabular}{cccccc}
    \toprule
         &  SC-SIS & PC-Screen & Bcor-SIS & DC-SIS & MrDc-SIS \\
         \cmidrule(lr){2-2} \cmidrule(lr){3-3} \cmidrule(lr){4-4} \cmidrule(lr){5-5} \cmidrule(lr){6-6}
       $\mathcal{S}$.mean  & 3776.08  & 2441.05  & 3164.50  & 2852.32   & \textbf{}{401.94}  \\
       $\mathcal{S}$.std & 1010.82 & 1462.06 & 1399.92 & 1404.24 & 792.86 \\
       \bottomrule
    \end{tabular}
\end{table}
\begin{table}[H]
\centering
\caption{$\mathcal{P}_s$ and $\mathcal{P}_a$ under different thresholds for case (1) of Example 3, $d_1 = [n/\log(n)] = 37$, $d_2 = 2d_1$, and $d_3 = 3d_1$. If we choose top $d_3$ predictors, MrDc-SIS can select all true signals in 63\% of the time, while other four approaches can only select all true signals in less than 3\% of the time.}
\label{tblpareto12}
\scriptsize

\begin{tabular}{*{16}{c}}
  \toprule
  \multirow{2}{*}{Model Size} &  \multicolumn{5}{c}{SC-SIS}  & \multicolumn{5}{c}{PC-Screen} & \multicolumn{5}{c}{BCor-SIS}   \\
  \cmidrule(lr){2-6} \cmidrule(lr){7-11} \cmidrule(lr){12-16}
   &   \multicolumn{4}{c} {$\mathcal{P}_s$} & $\mathcal{P}_a$ & \multicolumn{4}{c} {$\mathcal{P}_s$} & $\mathcal{P}_a$ & \multicolumn{4}{c} {$\mathcal{P}_s$} & $\mathcal{P}_a$  \\
   \cmidrule(lr){2-5} \cmidrule(lr){6-6} \cmidrule(lr){7-10} \cmidrule(lr){11-11} \cmidrule(lr){12-15} \cmidrule(lr){16-16}
  & $X_2$ & $X_3$ & $X_{101}$ & $X_{102}$ & All & $X_2$ & $X_3$ & $X_{101}$ & $X_{102}$ & All & $X_2$ & $X_3$ & $X_{101}$ & $X_{102}$ & All  \\
  $d_1$ & 0.02  & 0.03  & 0.02  & 0.025 & 0.01 & 0.02  & 0.015 & 1    & 1 & 0.015 & 0.015 & 0.015 & 0.995 & 1 & 0.01  \\
  $d_2$ & 0.025 & 0.04  & 0.025 & 0.035 & 0.01 & 0.025 & 0.02  & 1    & 1 & 0.015 & 0.025 & 0.02  & 0.995 & 1 & 0.01  \\
  $d_3$ & 0.045 & 0.06  & 0.03  & 0.04 & 0.01 & 0.04  & 0.035 & 1    & 1 & 0.025 & 0.03  & 0.035 & 0.995 & 1 & 0.015  \\
  \midrule
  \multirow{2}{*}{Model Size} & \multicolumn{5}{c}{DC-SIS} & \multicolumn{5}{c}{MrDc-SIS} & & & & & \\
  \cmidrule(lr){2-6} \cmidrule(lr){7-11} 
  & \multicolumn{4}{c} {$\mathcal{P}_s$} & $\mathcal{P}_a$ & \multicolumn{4}{c} {$\mathcal{P}_s$} & $\mathcal{P}_a$ & & & & & \\
  \cmidrule(lr){2-5} \cmidrule(lr){6-6} \cmidrule(lr){7-10} \cmidrule(lr){11-11} 
  & $X_2$ & $X_3$ & $X_{101}$ & $X_{102}$ & All & $X_2$ & $X_3$ & $X_{101}$ & $X_{102}$ & All & & & & & \\
  $d_1$ & 0.015 & 0.015 & 0.995 & 1 & 0.01 & 0.605 & 0.64  & 1    & 1 & 0.56 & & & & & \\
  $d_2$ & 0.02 & 0.02 & 1   & 1 & 0.015 & 0.66 & 0.69 & 1   & 1 & 0.605 & & & & & \\
  $d_3$ & 0.02  & 0.025 & 1    & 1 & 0.015 & 0.69  & 0.735 & 1    & 1 & \textbf{0.63} & & & & & \\
  \bottomrule
\end{tabular}

\end{table}

Tables \ref{tblpareto11}, \ref{tblpareto12} and Fig.~\ref{figpareto1} demonstrate the results of all the five approaches in case (1) of Example 3. We can see that MrDc-SIS performs better than all the other four approaches under this messy setting with heavy tail and right skewness, being achieved an average $\mathcal{S}$ of 400. The average $\mathcal{S}$ of PC-Screen is 2400, and $\mathcal{S}$'s of DC-SIS, BCor-SIS and SC-SIS are even higher. The boxplots of $\mathcal{S}$'s in Fig.~\ref{figpareto1} demonstrates that the other four feature screening methods may not distinguish the truth from the noise, while MrDc-SIS is still robust to effectively select the true predictors. From table~\ref{tblpareto12} we can see that MrDc-SIS has a simultaneous rate of 0.63 under threshold $d_3$, while the rates of the other four methods are below 0.1 under the same thresholds. 

\begin{table}[H]
    \centering
    \caption{The minimum selection size $\mathcal{S}$ of each screening method for case (2) of Example 3 where we have mixed categorical and continuous responses. MrDc-SIS still works better than other approaches, it ranks all true signals to top 600, while others rank the true signals to 3000.}
    \label{tblpareto21}
    \begin{tabular}{cccccc}
    \toprule
         &  SC-SIS & PC-Screen & Bcor-SIS & DC-SIS & MrDc-SIS \\
         \cmidrule(lr){2-2} \cmidrule(lr){3-3} \cmidrule(lr){4-4} \cmidrule(lr){5-5} \cmidrule(lr){6-6}
       $\mathcal{S}$.mean  & 3671.39  & 2316.39  & 3017.87  & 2796.06   & \textbf{}{631.01}  \\
       $\mathcal{S}$.std & 1074.09 & 1447.70 & 1444.25 & 1424.84 & 1194.76 \\
       \bottomrule
    \end{tabular}
\end{table}

\begin{table}[H]
\centering
\caption{$\mathcal{P}_s$ and $\mathcal{P}_a$ under different thresholds for case (2) of Example 3, $d_1 = [n/\log(n)] = 37$, $d_2 = 2d_1$, and $d_3 = 3d_1$. If we choose top $d_3$ predictors, MrDc-SIS can select all true signals in 60\% of the time, while other four approaches can only select all true signals in less than 8\% of the time.}
\label{tblpareto22}
\scriptsize

\begin{tabular}{*{16}{c}}
  \toprule
  \multirow{2}{*}{Model Size} &  \multicolumn{5}{c}{SC-SIS}  & \multicolumn{5}{c}{PC-Screen} & \multicolumn{5}{c}{BCor-SIS}   \\
  \cmidrule(lr){2-6} \cmidrule(lr){7-11} \cmidrule(lr){12-16}
   &   \multicolumn{4}{c} {$\mathcal{P}_s$} & $\mathcal{P}_a$ & \multicolumn{4}{c} {$\mathcal{P}_s$} & $\mathcal{P}_a$ & \multicolumn{4}{c} {$\mathcal{P}_s$} & $\mathcal{P}_a$  \\
   \cmidrule(lr){2-5} \cmidrule(lr){6-6} \cmidrule(lr){7-10} \cmidrule(lr){11-11} \cmidrule(lr){12-15} \cmidrule(lr){16-16}
  & $X_2$ & $X_3$ & $X_{101}$ & $X_{102}$ & All & $X_2$ & $X_3$ & $X_{101}$ & $X_{102}$ & All & $X_2$ & $X_3$ & $X_{101}$ & $X_{102}$ & All  \\
  $d_1$ & 0.035 & 0.04  & 0.025 & 0.05 & 0.01 & 0.04  & 0.045 & 0.995 & 1 & 0.03 & 0.035 & 0.035 & 0.99  & 1 & 0.025  \\
  $d_2$ & 0.045 & 0.05  & 0.03  & 0.09 & 0.01 & 0.08  & 0.065 & 1  & 1 & 0.055 & 0.04  & 0.035 & 0.99  & 1 & 0.025  \\
  $d_3$ & 0.065 & 0.07  & 0.035 & 0.11 & 0.01 & 0.095 & 0.09  & 1  & 1 & 0.08 & 0.045 & 0.04  & 0.99  & 1 & 0.025  \\
  \midrule
  \multirow{2}{*}{Model Size}  & \multicolumn{5}{c}{DC-SIS} & \multicolumn{5}{c}{MRDC-SIS} & & & & & \\
  \cmidrule(lr){2-6} \cmidrule(lr){7-11} 
  & \multicolumn{4}{c} {$\mathcal{P}_s$} & $\mathcal{P}_a$ & \multicolumn{4}{c} {$\mathcal{P}_s$} & $\mathcal{P}_a$ & & & & & \\
  \cmidrule(lr){2-5} \cmidrule(lr){6-6} \cmidrule(lr){7-10} \cmidrule(lr){11-11} 
  & $X_2$ & $X_3$ & $X_{101}$ & $X_{102}$ & All & $X_2$ & $X_3$ & $X_{101}$ & $X_{102}$ & All & & & & & \\
  $d_1$ & 0.035 & 0.035 & 0.995 & 1 & 0.03 & 0.61  & 0.585 & 1  & 1 & 0.54 & & & & & \\
  $d_2$ & 0.045 & 0.04  & 0.995 & 1 & 0.035 & 0.67  & 0.665 & 1  & 1 & 0.605 & & & & & \\
  $d_3$ & 0.045 & 0.04  & 0.995 & 1 & 0.035 & 0.68  & 0.675 & 1   & 1 & \textbf{0.605} & & & & & \\
  \bottomrule
\end{tabular}

\end{table}

Tables \ref{tblpareto21}, \ref{tblpareto22} and Fig.~\ref{figpareto2} demonstrate the results of all the five approaches for case (2) of Example 3. We can see that MrDc-SIS can still outperform the other four screening methods with an average $\mathcal{S}$ of 600 when both predictors and responses mixed with discrete and continuous components, while the other four approaches are trapped and need average $\mathcal{S}$'s more than 2000 to select true predictors. From Table~\ref{tblpareto22} we can see that MrDc-SIS has a simultaneous success rate $\mathcal{P}_a$ 0.6 under threshold $d_3$, while the rates of other four methods are still below 0.1 under the same threshold. 

\section{Real data analysis}
In this section we study cancer genome TCGA-OV data. The data is downloaded by TCGA-Assembler 2 with $\textsf{R}$ \citep{wei, tcgaass}. Altogether, we analyze data collected from three platforms: DNA copy number variation (CNV), DNA methylation (ME), and gene expression (GE) for each of the 296 patients. The modeling aim is to detect important associations between copy number variation, methylation, and gene expression, and hence detect active genes influencing the OV disease. 

\subsection{Data preprocessing}
Before the analysis, we preprocess the data as follows:
\begin{itemize}
  \item ME. We download the \emph{methylation\_27k} data. After removing missing values,  the DNA methylation data is arranged in a $296\times 14,280$ matrix.
  \item CNV. We download the \emph{nocnv.hg19} data, which eliminates some germline CNV. After removing missing values, the CNV data is arranged in a $296\times 13,956$ matrix.
  \item Integrated predictor array. We integrate CNV and ME in one unit. After matching the common genes of these two platforms, it results in a $ 296\times 13,491\times 2$ array, where the sample size is 296, the number of predictors is 13,491, and each predictor is a 2-dimensional vector (CNV for the 1st dimension and ME for the 2nd dimension). 
   \item GE. The expression values of five genes, \emph{BRCA1, BRCA2, TP53, BCL2L1, KRAS}, are the response and arranged in a $296\times 5$ matrix. We found these five genes from an overview link  \url{http://www.cancerindex.org/geneweb/X1003.htm}. This overview summarizes and ranks genes based on the number of findings that they were reported by thousands of publications with the relevant research and, it highlights the importance of these five genes in influencing ovarian cancer.
\end{itemize}

\begin{figure}[H]
\includegraphics[width=0.48\textwidth]{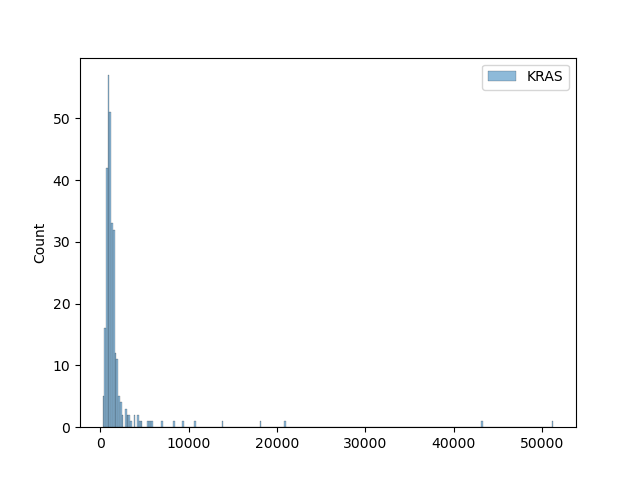}
\includegraphics[width=0.48\textwidth]{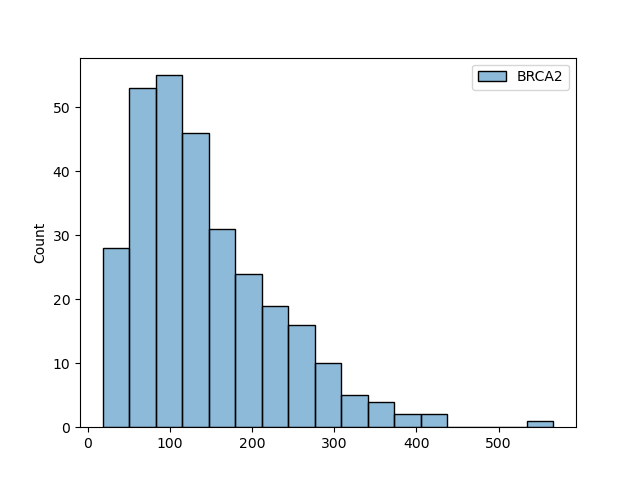}
\vspace{-6mm}
\caption{The histogram of gene expression data for gene \emph{KRAS} (left panel) and gene \emph{BRCA2} (right panel), as examples.}
\label{fig17}
\end{figure}

The explorative visualizations reveal some of the messy aspects of the dataset. For example, on the left panel of Fig.~\ref{fig17}, we can see an extremely long tail and large range (from 0 to 50000) for the expression values of gene \emph{KRAS}; on the right panel, heavy right skewness and a likely violation of the normality distribution assumption are demonstrated for the expression values of gene \emph{BRCA2}. The simulation studies already illustrated how challenging these issues are for existing feature screening approaches, which confirms the motivation of proposing the MrDc-SIS approach.

\subsection{Feature screening for integrated CNV and Methylation}
It has multiple advantages to treat the CNV and ME as a 2-dimensional unit rather than model each of them individually and separately: the predictor size will always be $p$ no matter how many different platforms we have, and it will be effortless to add new data; the correlation of multiple platforms of the same gene can be incorporated. As a multivariate feature screening approach, we want to evaluate the association strength between the 5-dimensional response vector and each of the 2-dimensional predictor vectors.

\begin{figure}[H]
\centering
\includegraphics[width=1.0\textwidth]{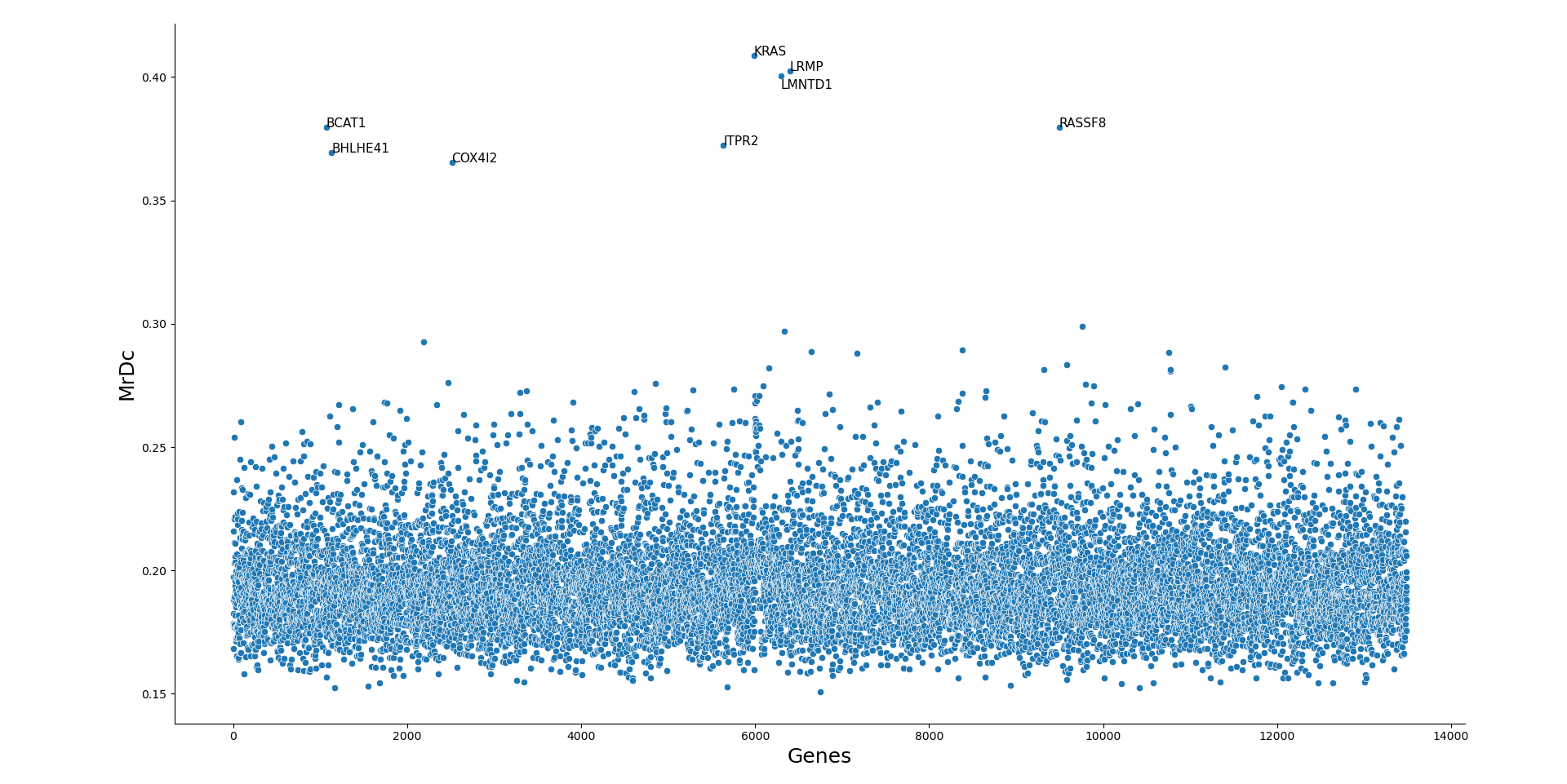}
\vspace{-9mm}
\caption{The multivariate rank distance correlation scores between the 5-dimensional response vector and each of the 13,491 2-dimensional predictors.}
\label{fig13}
\end{figure}
The MrDc scores of the 13,491 genes are demonstrated in Fig. \ref{fig13}. We can see that eight genes stand out and their MrDc scores are far beyond those of the remaining genes. We apply the max-ratio rule on these 13,491 scores and it chooses only the top eight genes as the important subset. They are \emph{BCAT1, BHLHE41, COX4I2, ITPR2, KRAS, LMNTD1, LRMP} and \emph{RASSF8}. Some of these findings quantitatively confirm other reports of the literature. For example, \emph{KRAS} and \emph{BRCA1} were identified to be associated with significant increased risk of developing ovarian cancer \citep{ratner2010kras}. The \emph{COX4I2} was identified as a molecular marker of the abnormal energy metabolism of cancer tissues by \citet{li2019signaling}. The \emph{LRMP} was identified to be amplified and over-expressed in ovarian cancer cell  \citep{tsuda2004identification}.

\begin{figure}[H]
\centering
\includegraphics[width=0.3\textwidth]{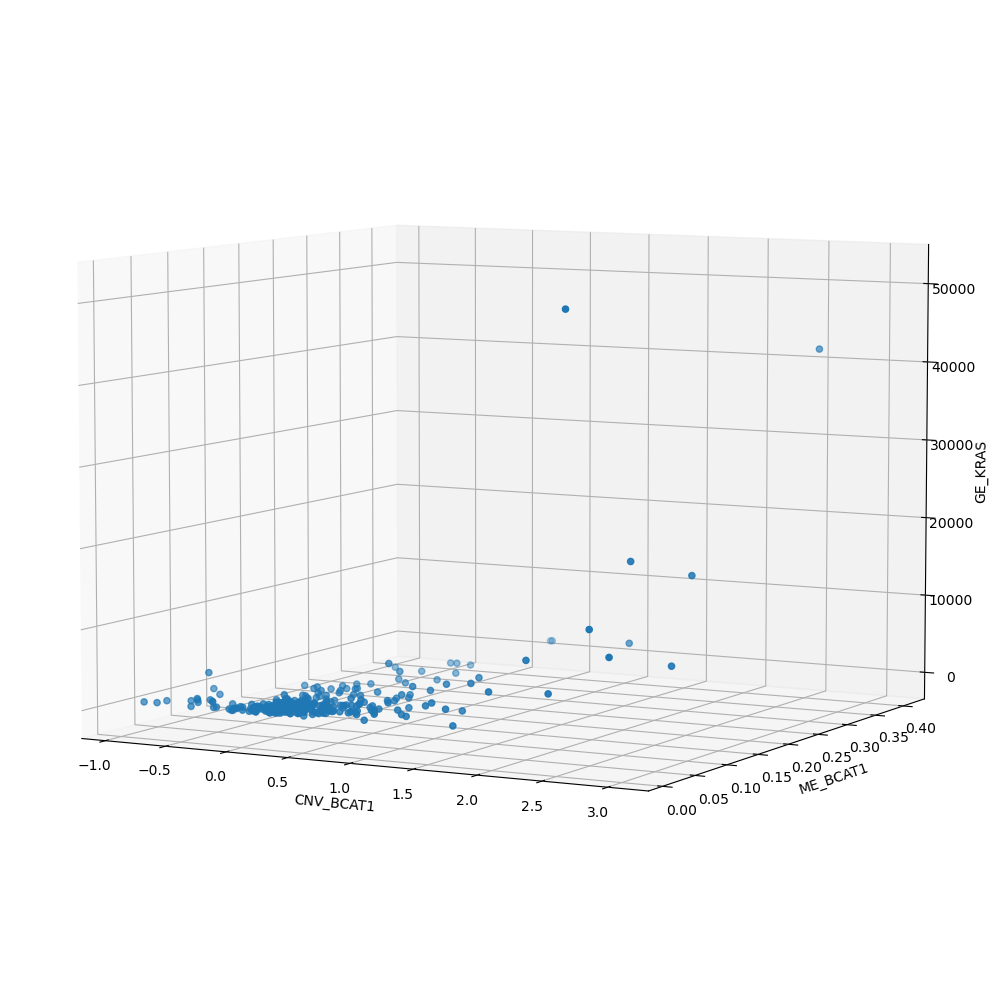}
\includegraphics[width=0.3\textwidth]{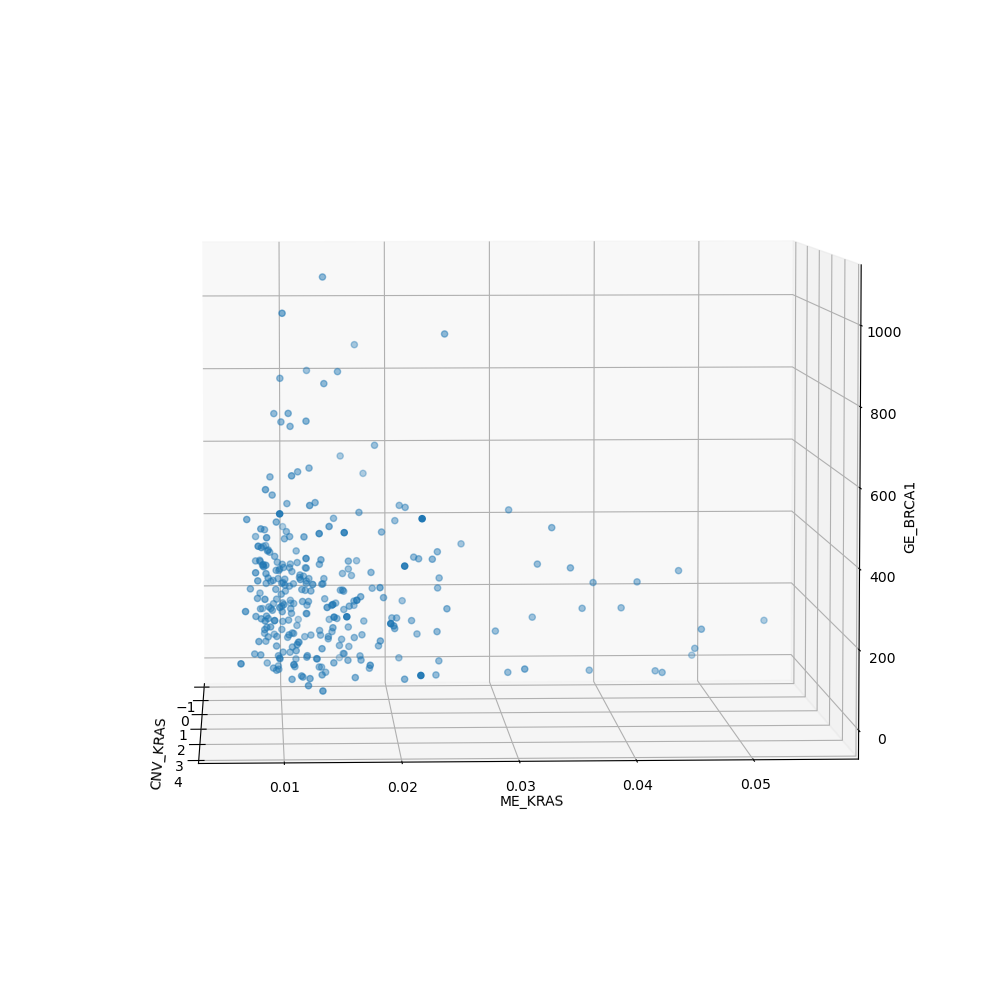}
\includegraphics[width=0.3\textwidth]{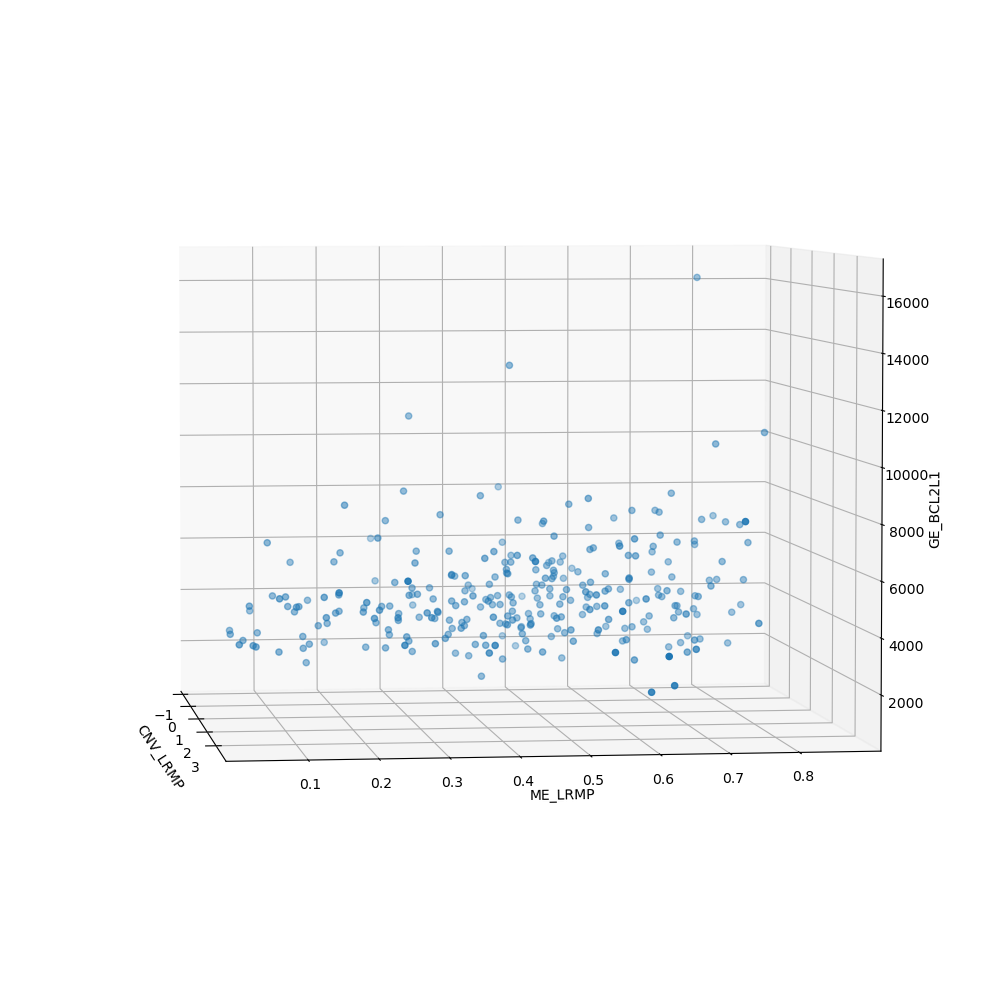}
\vspace{-9mm}
\caption{Visualization of some complex relationships. Left: GE of \emph{KRAS} versus the CNV and ME of \emph{BCAT1}; Middle: GE of \emph{BRCA1} versus the CNV and ME of \emph{KRAS}; and Right: GE of \emph{BCL2L1} versus the CNV and ME of \emph{LRMP}.}
\label{fig14}
\end{figure}

We also visually explore the 3D scatter plots of these important findings and notice some complex relationships. For example, a nonlinear trend between the CNV of \emph{BCAT1} with expression of gene \emph{KRAS} (left panel of Fig.~\ref{fig14}); a L-shaped trend between the Methylation of \emph{KRAS} and expression of gene \emph{BRCA1} (middle panel of Fig.~\ref{fig14}); and a megaphone-shaped trend between the Methylation of \emph{LRMP} and expression of gene \emph{BCL2L1} (right panel of Fig.~\ref{fig14}).  Fig.~\ref{fig14} not only illustrates that the proposed MrDc-SIS approach effectively captures complex relationships that are usually need to be solved by data transformation when using traditional approaches, but also demonstrates that multiple genes act together to influence the ovarian cancer cells. 

\section{Discussion}

In this paper, we propose a sure independence screening procedure via multivariate rank distance correlation. The main contribution of MrDc-SIS to the feature screening literature is that it provides an alternative robust feature screening approach that can be applied to multivariate predictors and multivariate responses. Without tail bound (sub-exponential or sub-Gaussian) restrictions for both the predictor and response, we prove the theoretical sure screening and rank consistency properties when the number of predictors diverges at an exponential rate of the sample size, which requires a minimal condition. Extensive simulation studies with various difficulty levels demonstrate that the proposed MrDc-SIS is robust and can capture both linear and nonlinear dependence structures and also mixed continuous and discrete data types.

\ta{In addition to the ``low-discrepancy" property, the low-discrepancy sequences that we utilized have more advantages. For example, if one extra observation is added, generating low-discrepancy sequences simply requires to add one more point without the need of recalculating the entire sequence, which greatly reduces the computational cost}. Furthermore, low-discrepancy sequences are fixed sequences and it leads to a reliable and reproducible consequence. 

We apply the MrDc-SIS approach to the multi-omics TCGA-OV dataset to detect important associations between gene expression, DNA methylation, and CNV. Eight genes, \emph{BCAT1, BHLHE41, COX4I2, ITPR2, KRAS, LMNTD1, LRMP} and \emph{RASSF8}, are detected to be important in associating with the ovarian cancer. Some of these results quantitatively confirm the findings of the literature that have also reported \emph{KRAS}, \emph{BRCA1} \emph{COX4I2}, \emph{LRMP} for OV disease using dramatically different methods and/or datasets \citep{ratner2010kras, tsuda2004identification,li2019signaling}.

The existing feature screening approaches count on a threshold to determine the selection size without computing a p-value. In this article, we utilize max-ratio criterion to determine the threshold for the selection size. Actually, the low discrepancy sequence is a fixed value sequence, one can easily get all the possible permutations of the sequence before even having the dataset. It makes the derivations of the distributions and p-value quick and easy without the need of permuting each dataset at a large number of times, which greatly reduces the computational cost if the number of predictors is large. Determining a p-value is important to access the significance of each predictor in biomedical applications. In the future work, we will explore p-values of the predictors under various dimensions of the dataset, which will greatly expand the application scope of the MrDc-SIS approach.

\section*{Acknowledgments}

We thank the Editor, Associate Editor and three anonymous referees for their valuable time and constructive comments.

\section*{Appendix}
\begin{lemma} (\citet[Lemma A, pp. 200]{serf}), Let $\mu=E(Y)$. If $\Pr(a\leqslant Y\leqslant b)=1$, then
\begin{equation*}
E[\exp\{s(Y-\mu)\}]\leqslant \exp\{s^2(b-a)^2/8\}, \ for\ any\ s>0.
\end{equation*}
\end{lemma}

\begin{lemma} (\citet[Theorem A, pp. 201]{serf}), Let $h(X_1,\dots, X_m)$ be a kernel of the U-statistic $U_n$, and $\theta =E\{h(X_1,\dots,X_m)\}$. If $a\leqslant h(X_1\dots,X_m)\leqslant b$, then for any $t>0$ and $n\leqslant m$,
\begin{equation*}
\Pr(U_n-\theta\geqslant t)\leqslant \exp \{-2[n/m]t^2/(b-a)^2\},
\end{equation*}
where $[n/m]$ denotes the integer part of $n/m$. Because of the symmetry of U-statistic, we have
\begin{equation*}
\Pr(|U_n-\theta|\geqslant t)\leqslant 2\exp \{-2[n/m]t^2/(b-a)^2\}.
\end{equation*}
\end{lemma}

\begin{proof}[\textbf {\upshape{Proof of (\ref{eqnew1}) in the Theorem 1:}}] We want to show the uniform consistency of the denominator and the numerator of $\hat{\omega}_j$, since the denominator and numerator have similar form, we only deal with the numerator. Define
\begin{equation*}
\begin{split}
\hat{S}_{j1}:=& \frac{1}{n^2}\sum_{k,l=1}^n\Vert\hat{R}_n^\Xbf(\Xbf_{jk})-\hat{R}_n^\Xbf(\Xbf_{j\ell})\Vert_d\Vert\hat{R}_n^\Ybf(\Ybf_{k})-\hat{R}_n^\Ybf(\Ybf_{\ell})\Vert_q, \\
\hat{S}_{j2}:=&(\frac{1}{n^2}\sum_{k,l=1}^n\Vert\hat{R}_n^\Xbf(\Xbf_{jk})-\hat{R}_n^\Xbf(\Xbf_{j\ell})\Vert_d)\times(\frac{1}{n^2}\sum_{k,l=1}^n\Vert\hat{R}_n^\Ybf(\Ybf_{k})-\hat{R}_n^\Ybf(\Ybf_{\ell})\Vert_q),\\
\hat{S}_{j3}:=&\frac{1}{n^3}\sum_{k,l,m=1}^n\Vert\hat{R}_n^\Xbf(\Xbf_{jk})-\hat{R}_n^\Xbf(\Xbf_{j\ell})\Vert_d\Vert\hat{R}_n^\Ybf(\Ybf_{k})-\hat{R}_n^\Ybf(\Ybf_{m})\Vert_q,
\end{split}
\end{equation*}
where $d$, $q$ stand for the dimension of $\Xbf_j$ and $\Ybf$, most of the time $d = 1$, but sometimes when the data are from different sources or in multiomics data, we may have $d>1$. Moreover,
\begin{equation*}
\begin{split}
S_{j1}=& E[\Vert R^\Xbf(\Xbf_j^1)-R^\Xbf(\Xbf_j^2)\Vert_d\Vert R^\Ybf(\Ybf^1)-R^\Ybf(\Ybf^2)\Vert_q], \\
S_{j2}=& E[\Vert R^\Xbf(\Xbf_j^1)-R^\Xbf(\Xbf_j^2)\Vert_d]E[\Vert R^\Ybf(\Ybf^1)-R^\Ybf(\Ybf^2)\Vert_q], \\
S_{j3}=& E[\Vert R^\Xbf(\Xbf_j^1)-R^\Xbf(\Xbf_j^2)\Vert_d\Vert R^\Ybf(\Ybf^1)-R^\Ybf(\Ybf^3)\Vert_q],
\end{split}
\end{equation*}
where $(\Xbf_j^1,\Ybf^1),(\Xbf_j^2,\Ybf^2),(\Xbf_j^3,\Ybf^3)$ are independent observations having the same distribution as $(\Xbf_j,\Ybf)$.\\
Firstly we focus on $\hat{S}_{j\ell}$, define
\begin{equation*}
\hat{S}_{j1}^*:=\frac{1}{n(n-1)}\sum_{k\neq \ell}\Vert\hat{R}_n^\Xbf(\Xbf_{jk})-\hat{R}_n^\Xbf(\Xbf_{j\ell})\Vert_d\Vert\hat{R}_n^\Ybf(\Ybf_{k})-\hat{R}_n^\Ybf(\Ybf_{\ell})\Vert_q.
\end{equation*}
By the triangle inequality, for any $k,\ell\in\{1,2,\dots,n\}$:
\begin{equation}
  \Vert\hat{R}_n^\Xbf(\Xbf_{jk})-\hat{R}_n^\Xbf(\Xbf_{j\ell})\Vert_d\geqslant\Vert R^\Xbf(\Xbf_{jk})-R^\Xbf(\Xbf_{j\ell})\Vert_d-\Vert\hat{R}_n^\Xbf(\Xbf_{jk})-R^\Xbf(\Xbf_{jk})\Vert_d-\Vert \hat{R}_n^\Xbf(\Xbf_{j\ell})-R^\Xbf(\Xbf_{j\ell})\Vert_d.
\end{equation}
So we have
\begin{equation}
\begin{split}
  \liminf_{n\to\infty}\hat{S}_{j1}^*\geqslant &\liminf_{n\to\infty}(\frac{1}{n(n-1)}\sum_{k\neq l}\Vert R^\Xbf(\Xbf_{jk})-R^\Xbf(\Xbf_{j\ell})\Vert_d\Vert\hat{R}_n^\Ybf(\Ybf_k)-\hat{R}_n^\Ybf(\Ybf_{\ell})\Vert_q) \\
  & -\limsup_{n\to\infty}(\frac{1}{n(n-1)}\sum_{k\neq l}\Vert\hat{R}_n^\Xbf(\Xbf_{jk})-R^\Xbf(\Xbf_{jk})\Vert_d\Vert\hat{R}_n^\Ybf(\Ybf_k)-\hat{R}_n^\Ybf(\Ybf_{\ell})\Vert_q) \\
  & -\limsup_{n\to\infty}(\frac{1}{n(n-1)}\sum_{k\neq l}\Vert \hat{R}_n^\Xbf(\Xbf_{j\ell})-R^\Xbf(\Xbf_{j\ell})\Vert_d\Vert\hat{R}_n^\Ybf(\Ybf_k)-\hat{R}_n^\Ybf(\Ybf_{\ell})\Vert_q).
\end{split}
\end{equation}
By Theorem 2.1 in Deb and Sen\cite{rdc}, the last two terms equal to 0 $a.s.$. Therefore
\begin{equation}
  \liminf_{n\to\infty}\hat{S}_{j1}^*\geqslant \liminf_{n\to\infty}(\frac{1}{n(n-1)}\sum_{k\neq l}\Vert R^\Xbf(\Xbf_{jk})-R^\Xbf(\Xbf_{j\ell})\Vert_d\Vert\hat{R}_n^\Ybf(\Ybf_k)-\hat{R}_n^\Ybf(\Ybf_{\ell})\Vert_q) \ \ a.s..
\end{equation}
Repeat the same argument on $\Ybf$'s instead of $\Xbf_j$'s, we can get
\begin{equation}
  \liminf_{n\to\infty}\hat{S}_{j1}^*\geqslant \liminf_{n\to\infty}(\frac{1}{n(n-1)}\sum_{k\neq l}\Vert R^\Xbf(\Xbf_{jk})-R^\Xbf(\Xbf_{j\ell})\Vert_d\Vert R^\Ybf(\Ybf_k)-R^\Ybf(\Ybf_{\ell})\Vert_q) \ \ a.s..
\end{equation}
Another application of the triangle inequality also yields the following:
\begin{equation}
  \Vert\hat{R}_n^\Xbf(\Xbf_{jk})-\hat{R}_n^\Xbf(\Xbf_{j\ell})\Vert_d\leqslant \Vert R^\Xbf(\Xbf_{jk})-R^\Xbf(\Xbf_{j\ell})\Vert_d+\Vert\hat{R}_n^\Xbf(\Xbf_{jk})-R^\Xbf(\Xbf_{jk})\Vert_d+\Vert \hat{R}_n^\Xbf(\Xbf_{j\ell})-R^\Xbf(\Xbf_{j\ell})\Vert_d.
\end{equation}
Using the similar arguments as before, we can get
\begin{equation}
  \limsup_{n\to\infty}\hat{S}_{j1}^*\leqslant \limsup_{n\to\infty}(\frac{1}{n(n-1)}\sum_{k\neq l}\Vert R^\Xbf(\Xbf_{jk})-R^\Xbf(\Xbf_{j\ell})\Vert_d\Vert R^\Ybf(\Ybf_k)-R^\Ybf(\Ybf_{\ell})\Vert_q) \ \ a.s..
\end{equation}
Define
\begin{equation}\label{eq1}
S_{j1}^{**}=\frac{1}{n(n-1)}\sum_{k\neq l}\Vert R^\Xbf(\Xbf_{jk})-R^\Xbf(\Xbf_{j\ell})\Vert_d\Vert R^\Ybf(\Ybf_k)-R^\Ybf(\Ybf_{\ell})\Vert_q.
\end{equation}
By the Cauchy-Schwartz inequality,
\begin{equation}
\begin{split}
  S_{j1}& = E[\Vert R^\Xbf(\Xbf_j^1)-R^\Xbf(\Xbf_j^2)\Vert_d\Vert R^\Ybf(\Ybf^1)-R^\Ybf(\Ybf^2)\Vert_q] \leqslant \{E(\Vert R^\Xbf(\Xbf_j^1)-R^\Xbf(\Xbf_j^2\Vert_d^2)E(\Vert R^\Ybf(\Ybf^1)-R^\Ybf(\Ybf^2)\Vert_q^2)\}^{1/2} \\
  & \leqslant \{4E(\Vert R^\Xbf(\Xbf_j)\Vert_d^2)4E(\Vert R^\Ybf(\Ybf)\Vert_q^2)\}^{1/2} < \infty.
  \end{split}
\end{equation}
We know $R^\Xbf(\Xbf_j)$ and $R^\Ybf(\Ybf)$ are uniformly bounded. For any given $\epsilon>0$, take $n$ large enough such that $S_{j1}/n<\epsilon$ and $|S_{j1}^{**}-\hat{S}_{j1}^*|<\epsilon$, then
\begin{equation*}
\begin{split}
  \Pr\{|\hat{S}_{j1}-S_{j1}|\geqslant 3\epsilon\} &=\Pr\{|\hat{S}_{j1}^*\frac{n-1}{n}-S_{j1}\frac{n-1}{n}-\frac{S_{j1}}{n}|\geqslant 3\epsilon\} \leqslant \Pr\{|\hat{S}_{j1}^*-S_{j1}|\frac{n-1}{n}\geqslant 3\epsilon-\frac{S_{j1}}{n}\} \\
  &\leqslant \Pr\{|\hat{S}_{j1}^*-S_{j1}|\geqslant 2\epsilon\} \leqslant \Pr\{|S_{j1}^{**}-S_{j1}|\geqslant \epsilon\}.
  \end{split}
\end{equation*}
So to establish the uniform consistency of $\hat{S}_{j1}$, it suffices to show the uniform consistency of $S_{j1}^{**}$, Notice $\{R^\Xbf(\Xbf_{jk}),R^\Ybf(\Ybf_k)\}_{1\leqslant k\leqslant n}$ are i.i.d. random vectors, so the right hand side of~\ref{eq1} is a standard U-statistic. Let $h_1(R^\Xbf(\Xbf_{jk}),R^\Ybf(\Ybf_k);R^\Xbf(\Xbf_{j\ell}),R^\Ybf(\Ybf_{\ell}))=\Vert R^\Xbf(\Xbf_{jk})-R^\Xbf(\Xbf_{j\ell})\Vert_d\Vert R^\Ybf(\Ybf_k)-R^\Ybf(\Ybf_{\ell})\Vert_q$ be the kernel of U-statistic $S_{j1}^{**}$, then
\begin{equation*}
  \begin{split}
     h_1 & =\Vert R^\Xbf(\Xbf_{jk})-R^\Xbf(\Xbf_{j\ell})\Vert_d\Vert R^\Ybf(\Ybf_k)-R^\Ybf(\Ybf_{\ell})\Vert_q \leqslant (\Vert R^\Xbf(\Xbf_{jk})\Vert_d+\Vert R^\Xbf(\Xbf_{j\ell})\Vert_d)(\Vert R^\Ybf(\Ybf_k)\Vert_q+\Vert R^\Ybf(\Ybf_{\ell})\Vert_q) \\
       & \leqslant 4\sqrt{ndq}\doteq M.
  \end{split}
\end{equation*}
By the Markov inequality, for any $t>0$ we have
\begin{equation*}
\Pr(S_{j1}^{**}-S_{j1}\geqslant \epsilon)\leqslant \frac{E[\exp(tS_{j1}^{**})]}{\exp(t\epsilon+tS_{j1})}.
\end{equation*}
By Serfling (1980, section 5.1.6) \cite{serf}, any U-statistic can be represented as an average of averages of i.i.d. random variables. So $S_{j1}^{**}=\frac{1}{n!}\sum_{n!}\Omega(\Xbf_{j1},\Ybf_1;\dots, \Xbf_{jn},\Ybf_n)$, where $\sum_{n!}$ denotes the summation over all permutations of $\{1,\dots,n\}$, and $\Omega$ is an average of $m=[n/2]$ i.i.d. random variables, $\Omega=\frac{1}{m}\sum_r h_1^{(r)}$. Since the exponential function is convex, by Jensen's inequality.
\begin{equation}
  E[\exp(tS_{j1}^{**})]=E[\exp \{t\frac{1}{n!}\sum_{n!}\Omega(\Xbf_{j1},\Ybf_1;\dots, \Xbf_{jn},\Ybf_n)\}]\leqslant \frac{1}{n!}\sum_{n!}E[\exp\{t\Omega(\Xbf_{j1},\Ybf_1;\dots, \Xbf_{jn},\Ybf_n)\}] =E^m\{\exp (\frac{1}{m}th_1^{(r)})\}.
\end{equation}
Apply Lemma 1, we have
\begin{equation}
  \Pr(S_{j1}^{**}-S_{j1}\geqslant \epsilon)\leqslant \frac{E^m\{\exp(\frac{1}{m}t[h_1^{(r)}-S_{j1}])\}}{\exp(t\epsilon)} 
   \leqslant \exp\{-t\epsilon+M^2t^2/(8m)\}.
\end{equation}
If we choose $t=\frac{4\epsilon m}{M^2}$, and by symmetry of U-statistic, we have
\begin{equation}
  \Pr(|S_{j1}^{**}-S_{j1}|\geqslant \epsilon)\leqslant 2\exp (-2\epsilon^2m/M^2).
\end{equation}
If we choose $M=cn^\gamma$ for $0<\gamma<\frac{1}{2}-\kappa$, when $n$ is sufficiently large, we can further get
\begin{equation}\label{1}
  \Pr(|\hat{S}_{j1}-S_{j1}|\geqslant 3\epsilon)\leqslant 2\exp(-c\epsilon^2n^{1-2\gamma}).
\end{equation}
Next we come to $\hat{S}_{j2}\doteq \hat{S}_{j2,1}\times \hat{S}_{j2,2}$ where $\hat{S}_{j2,1}=\frac{1}{n^2}\sum_{k\neq l}\Vert\hat{R}_n^\Xbf(\Xbf_{jk})-\hat{R}_n^\Xbf(\Xbf_{j\ell})\Vert_d$, $\hat{S}_{j2,2}=\frac{1}{n^2}\sum_{k\neq l}\Vert\hat{R}_n^\Ybf(\Ybf_{k})-\hat{R}_n^\Ybf(\Ybf_{\ell})\Vert_q$. Correspondingly, we write $S_{j2}=S_{j2,1}\times S_{j2,2}$, where $S_{j2,1}=E[\Vert R^\Xbf(\Xbf_j^1)-R^\Xbf(\Xbf_j^2)\Vert_d]$ and $S_{j2,2}=E[\Vert R^\Ybf(\Ybf^1)-R^\Ybf(\Ybf^2)\Vert_q]$. Following the same arguments for proving (\ref{1}) we can show
\begin{equation}\label{2}
  \Pr (|\hat{S}_{j2,1}-S_{j2,1}|\geqslant 3\epsilon)\leqslant 2\exp(-c\epsilon^2 n^{1-2\gamma})
\end{equation}
and
\begin{equation}\label{3}
  \Pr (|\hat{S}_{j2,2}-S_{j2,2}|\geqslant 3\epsilon)\leqslant 2\exp(-c\epsilon^2 n^{1-2\gamma}).
\end{equation}
Like previously, $S_{j2,1}$ and $S_{j2,2}$ are uniformly bounded, so
\begin{equation}\label{4}
  \max_{1\leqslant j\leqslant p} \{S_{j2,1}, S_{j2,2}\}\leqslant C
\end{equation}
for some constant $C$. We have the following
\begin{equation*}
  \begin{split}
     \Pr(|(\hat{S}_{j2,1}-S_{j2,1})S_{j2,2}|\geqslant \epsilon) & \leqslant \Pr(|\hat{S}_{j2,1}-S_{j2,1}|\geqslant \frac{\epsilon}{C})\leqslant 2\exp(\frac{-c\epsilon^2n^{1-2\gamma}}{9C^2}), \\
     \Pr(|(\hat{S}_{j2,2}-S_{j2,2})S_{j2,1}|\geqslant \epsilon)  & \leqslant \Pr(|\hat{S}_{j2,2}-S_{j2,2}|\geqslant \frac{\epsilon}{C})\leqslant 2\exp(\frac{-c\epsilon^2n^{1-2\gamma}}{9C^2}),
  \end{split}
\end{equation*}
and
\begin{equation*}
     \Pr(|(\hat{S}_{j2,1}-S_{j2,1})(\hat{S}_{j2,2}-S_{j2,2})|\geqslant\epsilon)  \leqslant \Pr(|(\hat{S}_{j2,1}-S_{j2,1})|\geqslant \sqrt{\epsilon})+\Pr(|(\hat{S}_{j2,2}-S_{j2,2})|\geqslant \sqrt{\epsilon}) 
        \leqslant 4\exp(\frac{-\epsilon n^{1-2\gamma}}{9}).
\end{equation*}
By the Bonferroni inequality, we have
\begin{equation}\label{5}
  \begin{split}
     & \Pr(|\hat{S}_{j2}-S_{j2}|\geqslant 3\epsilon) = \Pr(|\hat{S}_{j2,1}\hat{S}_{j2,2}-S_{j2,1}S_{j2,2}|\geqslant 3\epsilon) \\
        \leqslant & \Pr(|(\hat{S}_{j2,1}-S_{j2,1})S_{j2,2}|\geqslant \epsilon)+\Pr(|(\hat{S}_{j2,2}-S_{j2,2})S_{j2,1}|\geqslant \epsilon)+ \Pr(|(\hat{S}_{j2,1}-S_{j2,1})(\hat{S}_{j2,2}-S_{j2,2})|\geqslant\epsilon) \\
        \leqslant & 8\exp(\frac{-c\epsilon^2n^{1-2\gamma}}{9C^2}).
  \end{split}
\end{equation}
The last inequality holds when $\epsilon$ is sufficiently small and $C$ is sufficiently large. For $\hat{S}_{j3}$, we use the third order U-statistic and follow similar arguments can get
\begin{equation*}
  \begin{split}
     & \hat{S}_{j3}^* =\frac{6}{n(n-1)(n-2)} \sum_{k<l<m}h_3(\hat{R}_n^\Xbf(\Xbf_{jk}),\hat{R}_n^\Ybf(\Ybf_k);\hat{R}_n^\Xbf(\Xbf_{j\ell}),
     \hat{R}_n^\Ybf(\Ybf_{\ell});\hat{R}_n^\Xbf(\Xbf_{jm}),\hat{R}_n^\Ybf(\Ybf_m)), \\
     & S_{j3}^{**}   = \frac{6}{n(n-1)(n-2)} \sum_{k<l<m}h_3(R^\Xbf(\Xbf_{jk}),R^\Ybf(\Ybf_k);R^\Xbf(\Xbf_{j\ell}),
     R^\Ybf(\Ybf_{\ell});R^\Xbf(\Xbf_{jm}),R^\Ybf(\Ybf_m)),\\
     & \Pr(|S_{j3}^{**}-S_{j3}|\geqslant \epsilon)   \leqslant 2\exp(-2\epsilon^2m'/M^2)=2\exp(-2c\epsilon^2n^{1-2\gamma}/3).
  \end{split}
\end{equation*}
Here $m'=[n/3]$ since it is a third order U-statistic. And $h_3(X_1,Y_1;X_2,Y_2;X_3,Y_3)=\Vert X_1-X_2\Vert_d\Vert Y_1-Y_3\Vert_q+\Vert X_1-X_2\Vert_d \Vert Y_2-Y_3\Vert_q +\Vert X_1-X_3\Vert_d\Vert Y_1-Y_2\Vert_q+\Vert X_1-X_3\Vert_d\Vert Y_2-Y_3\Vert_q+\Vert X_2-X_3\Vert_d\Vert Y_1-Y_2\Vert_q+\Vert X_2-X_3\Vert_d\Vert Y_1-Y_3\Vert_q$ is the kernel of U-statistic. By definition of $\hat{S}_{j3}$ we know
\begin{equation*}
  \hat{S}_{j3}=\frac{(n-1)(n-2)}{n^2}(\hat{S}_{j3}^*+\frac{1}{n-2}\hat{S}_{j1}^*).
\end{equation*}
Again $S_{j3}$ is finite, take $n$ large enough such that $\frac{3n-2}{n^2}S_{j3}\leqslant\epsilon$ and $\frac{n-1}{n^2}S_{j1}\leqslant\epsilon$, then
\begin{equation*}
  \begin{split}
     \Pr(|\hat{S}_{j3}-S_{j3}|\geqslant 6\epsilon) =& \Pr(|\frac{(n-1)(n-2)}{n^2}(\hat{S}_{j3}^*-S_{j3})-\frac{3n-2}{n^2}S_{j3}+\frac{n-1}{n^2}(\hat{S}_{j1}^*-S_{j1})+\frac{n-1}{n^2}S_{j1}|\geqslant 6\epsilon) \\
      \leqslant &  \Pr(|\hat{S}_{j3}^*-S_{j3}|\geqslant 2\epsilon)+\Pr(|\hat{S}_{j1}^*-S_{j1}|\geqslant 2\epsilon) \leqslant   \Pr(|S_{j3}^{**}-S_{j3}|\geqslant \epsilon)+\Pr(|S_{j1}^{**}-S_{j1}|\geqslant \epsilon) \\
      \leqslant &  4\exp(-2\epsilon^2n^{1-2\gamma}/3).
  \end{split}
\end{equation*}
Combining these 3 inequalities, we have
\begin{equation*}
  \begin{split}
        \Pr(|(\hat{S}_{j1}+\hat{S}_{j2}-2\hat{S}_{j3})-(S_{j1}+S_{j2}-2S_{j3})|\geqslant \epsilon) & \leqslant \Pr(|\hat{S}_{j1}-S_{j1}|\geqslant \frac{\epsilon}{4})+\Pr(|\hat{S}_{j2}-S_{j2}|\geqslant \frac{\epsilon}{4})+2\Pr(|\hat{S}_{j3}-S_{j3}|\geqslant \frac{\epsilon}{4}) \\
       & =O(\exp(-c_1\epsilon^2n^{1-2\gamma})).
  \end{split}
\end{equation*}
for some positive constant $c_1$. This is the convergence rate of the numerator of $\hat{\omega}_j$. Since the denominator has the same form of numerator, let $\epsilon=cn^{-\kappa}$ and $0<\kappa+\gamma<\frac{1}{2}$, we have
\begin{equation*}
     \Pr(\max_{1\leqslant j\leqslant p}|\hat{\omega}_j-\omega_j|\geqslant cn^{-\kappa})  \leqslant p\max_{1\leqslant j\leqslant p}\Pr(|\hat{\omega}_j-\omega_j|\geqslant cn^{-\kappa}) \leqslant O(p\exp(-c_1n^{1-2(\kappa+\gamma)}))
\end{equation*}
\end{proof}

\vspace{10mm}
\begin{proof}[\textbf {\upshape{ Proof of (\ref{eqnew2}) in the Theorem 1:}}] If $\mathcal{D}\nsubseteq \hat{\mathcal{D}}$, then there must exist some $j\in\mathcal{D}$ such that $\hat{\omega}_j<cn^{-\kappa}$, since $\omega_j\geqslant 2cn^{-\kappa}$ for all $j\in\mathcal{D}$, we know there exist some $j\in\mathcal{D}$ such that $|\hat{\omega}_j-\omega_j|>cn^{-\kappa}$. So $\{\mathcal{D}\nsubseteq\hat{\mathcal{D}}\}\subset\{|\hat{\omega}_j-\omega_j|>cn^{-\kappa},\ for\ some\ k\in\mathcal{D}\}$. Hence $\{\max_{j\in\mathcal{D}}|\hat{\omega}_j-\omega_j|\leqslant cn^{-\kappa}\}\subset\{\mathcal{D}\subseteq\hat{\mathcal{D}}\}$.
\begin{equation*}
  \begin{split}
     & \Pr(\mathcal{D}\subseteq\hat{\mathcal{D}})  \geqslant \Pr(\max_{j\in\mathcal{D}}|\hat{\omega}_j-\omega_j|\leqslant cn^{-\kappa}) =1-\Pr(\min_{j\in\mathcal{D}}|\hat{\omega}_j-\omega_j|\geqslant cn^{-\kappa})=1-s_n \Pr(|\hat{\omega}_j-\omega_j|\geqslant cn^{-\kappa}) \\
       \geqslant & 1-O(s_n\exp(-c_1n^{1-2(\kappa+\gamma)})).
  \end{split}
\end{equation*}
where $s_n$ is the cardinality of $\mathcal{D}$. This completes the proof.
\end{proof}

\begin{proof}[\textbf {\upshape{ Proof of Theorem 2}}]
We have 
\begin{equation*}
\begin{split}
\Pr(\min_{j\in\mathcal{D}}\hat\omega_j\leqslant \max_{j\in\mathcal{I}}\hat\omega_j) & = \Pr([\min_{j\in\mathcal{D}}\omega_j - \max_{j in \mathcal{I}}\omega_j] - [\min_{j\in\mathcal{D}}\hat\omega_j - \max_{j\in\mathcal{I}}\hat\omega_j]\geqslant 2c_2n^{-\kappa_2}) \\
& = \Pr([\max_{j\in\mathcal{I}}\hat\omega_j - \max_{j\in\mathcal{I}}\omega_j] - [\min_{j\in\mathcal{D}}\hat\omega_j - \min_{j\in\mathcal{D}}\omega_j]\geqslant 2c_2n^{-\kappa_2}) \\
& \leqslant \Pr(\max_{j\in\mathcal{I}}|\hat\omega_j - \omega_j| + \max_{j\in\mathcal{D}}| \hat\omega_j - \omega_j| \geqslant 2c_2n^{-\kappa_2}) \leqslant \Pr(\max_{1\leqslant j \leqslant p}|\hat{\omega}_j - \omega_j|\geqslant c_2n^{-\kappa_2}) \\
& \leqslant O(p\exp(c_1n^{1-2(\kappa_2+\gamma)})).
\end{split}
\end{equation*}
This completes the proof.
\end{proof}


\bibliographystyle{myjmva}
\bibliography{new}


\end{document}